\documentclass[10pt, 
               showpacs, 
         amsmath,amssymb,aps,prd,
              nofootinbib,eqsecnum, amsart,
                superscriptaddress,
               a4paper]{revtex4}

\usepackage{hyperref}
\usepackage{graphicx}
\usepackage{color}
\usepackage{bm}
\usepackage{mathtools}
\usepackage{wrapfig}
\usepackage{dcolumn}
\DeclareMathVersion{nxbold} 
\SetSymbolFont{operators}{nxbold}{OT1}{cmr} {b}{n}
\SetSymbolFont{letters}  {nxbold}{OML}{cmm} {b}{it}
\SetSymbolFont{symbols}  {nxbold}{OMS}{cmsy}{b}{n}

\usepackage{todonotes}

\input{_jnls.cls} 

\usepackage{siunitx}
\usepackage{adjustbox}
\usepackage{enumitem}
\usepackage{afterpage}
\usepackage{currfile}

\DeclareMathAlphabet{\mathpzc}{OT1}{pzc}{m}{it}

\newcommand{\infrac}[2]{#1/#2}
\newcommand{\arm}{a_\text{eq}}
\newcommand{\Hrm}{{\cal{H}}_\text{eq}}

\renewcommand{\vec}[1]{\mathbf{#1}}

\newcommand*\nc[1]{\tikz[baseline=(char.base)]{
    \node[shape=circle,draw,inner sep=2pt] (char) {#1};}}
\newcommand{\balg}{\begin{enumerate}[label=\protect\nc{\arabic*}]}
\newcommand{\ealg}{\vspace{6 pt}\end{enumerate}}

\newcommand{\smax}{s_\text{max}}

\newcommand{\tauc}{\tau_\text{c}}
\newcommand{\rhor}{\rho_\text{r}}
\newcommand{\rhom}{\rho_\text{m}}
\newcommand{\kunit}{\infrac{\Hrm}{\sqrt{2}}}
\newcommand{\tunit}{\infrac{\sqrt{2}}{\Hrm}}

\newcommand{\hmatch}{h_\text{rm}}

\newcommand{\hr}{h_\text{r}}
\newcommand{\tr}{\tau_\text{r}}

\ExplSyntaxOn
\NewDocumentCommand{\countlist}{m}
 {
  \clist_count:n { #1 }
 }
\ExplSyntaxOff
\newcommand{\rfcite}[1]{\ifnum\countlist{#1}=1  Ref.~\cite{#1}\else Refs.~\cite{#1}\fi}
\newcommand{\eref}[1]{Eq.~(\ref{#1})}

\renewcommand{\eqref}[1]{(\ref{#1})}

\DeclareMathOperator{\order}{\mathcal{O}}

\def\H{{\cal{H}}}
\def\bea{\begin{eqnarray}}
\def\eea{\end{eqnarray}}
\def\bi{\begin{itemize}}
\def\ei{\end{itemize}}
\def\be#1\ee{\begin{equation}#1\end{equation}}
\def\ba#1\ea{\begingroup
\addtolength{\jot}{9pt}\begin{align}#1\end{align}\endgroup}
\def\bas#1\eas{\begingroup
\addtolength{\jot}{9pt}\begin{align}\begin{split}#1\end{split}\end{align}\endgroup}
\def\bfa#1\efa{\begingroup
\addtolength{\jot}{9pt}\begin{flalign}#1\end{flalign}\endgroup}
\def\bml#1\eml{\begingroup
\addtolength{\jot}{9pt}\begin{multline}#1\end{multline}\endgroup}

\def\nt#1\ent{{\tt#1 }} 

\begin{document}

\makeatletter
\newcolumntype{Z}[3]{>{\mathversion{nxbold}\DC@{#1}{#2}{#3} }c<{\DC@end} } 
\makeatother
\newcommand{\bcell}[1]{\multicolumn{1}{Z{.}{.}{-1} }{#1}}

 
\title{Gravitational waves 
in a flat radiation--matter universe including anisotropic stress}

\author{Andrew~J.~Wren}
\email[]{andrew.wren@ntlworld.com}
\affiliation{Astronomy Unit, School of Physics and Astronomy, Queen Mary University of London, Mile End Road, London, E1 4NS, United Kingdom}
\author{Jorge~L.~Fuentes} \email[]{j.fuentesvenegas@qmul.ac.uk}
\affiliation{Astronomy Unit, School of Physics and Astronomy, Queen Mary University of London, Mile End Road, London, E1 4NS, United Kingdom}
\author{Karim~A.~Malik} \email[]{k.malik@qmul.ac.uk}
\affiliation{Astronomy Unit, School of Physics and Astronomy, Queen Mary University of London, Mile End Road, London, E1 4NS, United Kingdom}

\date{\today}

\begin{abstract}
We present novel analytical solutions for linear--order gravitational
waves or tensor perturbations in a flat Friedmann--Robertson--Walker
universe containing two perfect fluids, radiation and pressureless
dust, and allowing for neutrino anisotropic stress. One of the results
is applicable to any sub--horizon gravitational wave in such a
universe. Another result is applicable to gravitational waves of
primordial origin, for example produced during inflation, and works
both before and after they cross the horizon. These results improve on
analytical approximations previously set out in the
literature. Comparison with numerical solutions shows that both these
approximations are accurate to within $1\%$, or better, for a wide
range of wave--numbers relevant for cosmology.
\end{abstract}

\pacs{02.30.Hq, 02.30.Mv, 02.60.Gf, 04.25.Nx, 04.30.Tv, 95.35.+d, 95.30.Sf, 98.80.-k, 98.80.Cq,
98.80.Jk\hfill } 

\maketitle

\section{Introduction}
\label{sec:intro}

In 2016 observations by LIGO \cite{2016PhRvL.116f1102A} have
directly detected gravitational waves for the first time. The origin
of these waves was astrophysical: the merger of two large stellar mass
black holes.
In contrast, primordial gravitational waves originating from inflation
have been constrained, but not detected, by for example
\textsc{Planck} satellite observations, combined with data from
BICEP2/Keck Array \cite{PhysRevLett.114.101301} and with other data
sources \cite{2015arXiv150201589P}.  The simplest inflationary
models consistent with these results tend to favour inflationary
scenarios which generate gravitational waves of relatively low, but
non--negligible, amplitude compared with scalar perturbations.

In this paper we consider gravitational waves with sufficiently small
amplitude to be modelled as linear tensor perturbations travelling
through a flat Friedmann--Robertson--Walker (FRW) universe filled with
radiation and matter.  The matter component is pressureless dust, and our
perturbation equations treat both radiation and matter as perfect
fluids. We allow for a source term due to neutrino anisotropic stress.\\

As recalled in for example
\rfcite{1993PhRvD..48.4613T,1994PhRvD..50.3713A,1995PhRvD..52.2112N,1996PhRvD..53..639W,2004PhRvD..69b3503W,2005AnPhy.318....2P,2005astro.ph..5502B,2008cosm.book.....W,2013PhRvD..88h3536S},
there are well known exact analytical expressions for the evolution of
linear tensor perturbations in a flat universe dominated by either
radiation only or matter (pressureless dust) only. These
approximations do not work particularly well when used for a universe
containing both radiation and matter. The next simplest approximation
is to use the radiation--only solution before the time of
radiation--matter equality, and the matter--only solution after that
time. A slightly more sophisticated approach, used for example in
\rfcite{1994PhRvD..50.3713A,1995PhRvD..52.2112N}, is to match these
two solutions together at the time of radiation--matter equality,
termed the ``sudden approximation''.
Reference~\cite{1995PhRvD..52.2112N} shows that these approaches
do not provide particularly good approximations to a numerical
solution of the linear tensor perturbation equation.  In
\rfcite{1996PhRvD..53..639W}, a smoother transition is sought between
the radiation and matter solutions, fitting parameters to numerical
solutions, a method which is most suited to dealing with wave--numbers
much lower (corresponding to much larger wave--lengths) than those we
consider.

Later work in
\rfcite{2004PhRvD..69b3503W,2005AnPhy.318....2P,2005astro.ph..5502B,2008cosm.book.....W,2013PhRvD..88h3536S}
is aimed primarily approximating the effects of neutrino
free--streaming as expressed via an integro--differential equation
derived in \rfcite{PhysRevD.50.2541,2004PhRvD..69b3503W}, rather than at improving
accuracy in the perfect fluid model of radiation and matter.
In this paper we are improving the accuracy of the solutions also in the case of
non-zero anisotropic stress.\\

We now briefly survey the accuracy of the approximations of
\rfcite{1995PhRvD..52.2112N,2005AnPhy.318....2P,2005astro.ph..5502B,2008cosm.book.....W,2013PhRvD..88h3536S}. References~\cite{1995PhRvD..52.2112N,2005AnPhy.318....2P},
\rfcite{2008cosm.book.....W} and \rfcite{2005astro.ph..5502B} pursue
three different forms of the Wentzel-–Kramers-–Brillouin (WKB) method
for finding approximate analytical solutions for ordinary differential
equations.

The methods of \rfcite{1995PhRvD..52.2112N,2005AnPhy.318....2P} and  \rfcite{2008cosm.book.....W} provide results which, when restricted to the perfect fluid case, are accurate for primordial gravitational waves to within about $10\%$ for inverse wave--numbers, evaluated at the present day, of around \num{10}--\SI{15}{Mpc}. Reference~\cite{2005astro.ph..5502B} presents another form of WKB approximation for a flat radiation--matter universe. Restricted to the perfect fluid case, this provides a good sub--horizon approximation for gravitational waves of primordial origin. However, it does not retain such good accuracy outside the horizon, and covers only one of the two independent solutions of the governing equation for tensor perturbations. 

A different kind of approach is pursued in 
\rfcite{2013PhRvD..88h3536S}, using sums of spherical Bessel functions of successive orders to approximate tensor perturbations.  However, \rfcite{2013PhRvD..88h3536S}'s approach is best used for numerical calculations based on expansions of very high order~--- they exemplify solutions to 20th and 100th order.\\  

We derive improved results by employing a method set out, in
a non--cosmological context, by Feshchenko, Shkil' and Nikolenko (FSN)
in 1966 \cite{FSN}, and used by two of the authors in a recent paper \cite{DPS} to   approximate the solutions of the ordinary differential equations governing  scalar density perturbations. The FSN method is well
suited for solving linear second order ordinary
differential equations, that also depend on a small
parameter, which here we take to be the inverse wave--number.  In effect, this approach extends the WKB method of \rfcite{2008cosm.book.....W}.\\

Following this introduction, in
Section~\ref{sec:linear-tensor-perturbations} of this paper, we recall
the differential equation which governs tensor metric perturbations,
focusing on a flat radiation--matter universe, and also noting the
simpler differential equation for a flat radiation--only model and its
analytical solution with and without neutrino anisotropic stress.
In Section~\ref{sec:approximating-tensor-perturbations}, we then use
the method first set out in~\rfcite{FSN} to find an approximate
analytical solution to the governing equation of
Section~\ref{sec:linear-tensor-perturbations}.  Because tensor
perturbations are governed by only a single second order differential
equation, application of the method in this paper is more
straightforward than the work of \rfcite{DPS} on scalar perturbations
where a set of coupled differential equations is involved\footnote{The
  scalar perturbations of \rfcite{DPS} depend on a pair of second
  order differential equations to determine the radiation density and
  metric perturbations, plus an additional second order differential
  equation to get the matter density perturbation.}.

Our approximation method is based on the \emph{sub--horizon} assumption that the tensor perturbation has a wave--number larger than the Hubble parameter.  However, using numerical solutions, we can see that, for a wide range of wave--numbers, it is possible to extend the approximation back to earlier times when the wave--number is similar to the Hubble parameter. We find the approximation of third order in the inverse wave--number is accurate to within $1\%$ or better for gravitational waves with a present day inverse wave--number less than, or equal to, around  $\SI{17}{Mpc}.$ 
 
Section~\ref{sec:modelling-gravitational-waves-through-the-whole-radiation--matter-epoch} extends the analytical approximation to cover gravitational waves of primordial origin, from early times when their wave--length is larger than the horizon through their subsequent sub--horizon evolution. Our approximation to second order in the inverse wave--number accurate to within $1\%,$ or better,  for gravitational waves corresponding to those with an actual present day inverse wave--number less than, or equal to, around  $\SI{35}{Mpc}.$  This range includes wave--numbers which represent scales of key relevance for structure formation in the Universe, and broadly corresponds  to CMB multipoles $l\gtrsim 120$. 

Section~\ref{sec:conclusion} concludes the paper. It summarises the results and provides a brief discussion.

A \emph{Mathematica} notebook which executes the approximation method automatically is available online at \rfcite{GWGithub}.  It enables all the approximations presented in this paper to be calculated in less than a second on a standard PC. The template can be easily adapted for use with other, similar, second order differential equations.

Throughout this paper, we work in conformal time $\tau,$ with a dash
indicating the derivative of a function with respect to conformal
time.  We set the speed of light $c=1$, co--moving spatial
co-ordinates are denoted $x^i$ and Latin indices $i,j,k$ range from
$1$ to $3$.

\section{Linear tensor perturbations}
\label{sec:linear-tensor-perturbations}

We now study linear perturbations in a flat
Friedmann--Robertson--Walker (FRW) model with two non--interacting
perfect fluids~--- radiation and matter, the latter being pressureless
dust. We begin with the background equations, with the conformal time
Friedmann equation for such a model being given by
\be\label{eq:Friedmann}
\H^2=\frac{8\pi G}{3} a^2 \left(\rhor+\rhom \right)\,,
\ee
where $\H$ is the conformal Hubble factor, $G$ is the gravitational constant, $a$ is the scale factor, $\rhor$ the homogeneous radiation density, and $\rhom$ the homogeneous matter density. As usual, the radiation and matter densities obey
\ba \label{eq:densities}
\rhor\propto a^{-4} & \text{\qquad and\qquad }  \rhom\propto a^{-3}\,,
\ea
and, as
noted in, for example, \rfcite{DPS,2006AIPC..843..111P}, the scale factor is given by
\ba \label{eq:rmscalefactor}
a(\tau)=\arm\left(\frac{\tau}{\tauc}+\frac{\tau^2}{4\tauc^2}\right)\,,
\ea
where $\arm$ is the time of radiation--matter equality, $\tau$ is the conformal time, and $\tauc=\infrac{\sqrt{2}}{\Hrm}.$ In the following, we will use a normalised conformal time co--ordinate and replace $\infrac{\tau}{\tauc}$ by $\tau,$ giving us the simpler expression
\ba \label{eq:rmscale}
a(\tau)=\arm\left(\tau+\frac{\tau^2}{4}\right)
\,.
\ea
We also normalise the co--moving wave--number $k$ by using the corresponding inverse units $\tauc^{-1}=\infrac{\Hrm}{\sqrt{2}},$ and we note that, in those units, the conformal Hubble parameter is then given by
\ba\label{eq:h-in-tau}
\H(\tau)=\frac{2\left(2+\tau\right)}{\tau\left(4+\tau\right)}\,.
\ea

For future reference, using the cosmological parameters of \rfcite{2015arXiv150201589P}, we find that we have
\ba \label{eq:phys-unit} \kunit=\SI[group-digits = false]{0.00727}{a_0.Mpc^{-1}}
\,,
\ea
where $a_0$ is the value of the scale factor at the present time, \emph{assuming the universe contains only radiation and matter}.  To compare inverse wave--numbers with \emph{actual} present day distances, we use a value of $a_0$ which takes account of  the  expansion of the Universe due to dark energy.      From for example  \rfcite{2006PASP..118.1711W}, we find that the Universe is currently some $16\%$ bigger than it would have been without dark energy. If, as used implicitly in \rfcite{2015arXiv150201589P}, we take the actual present day value of $a_0$ to be $1,$ then, in the universe with only radiation and matter assumed in  \eref{eq:phys-unit} we have $a_0\approx\infrac{1}{1.16}\approx 0.86,$ giving us
\ba \label{eq:actual}
\kunit=\SI{0.0063}{Mpc^{-1}}\,.
\ea

For use below, in drawing figures directly comparable with those of \rfcite{2005AnPhy.318....2P}, we note the conformal time of recombination, $\tr.$  Using the redshift of $z_\star=\num{1089.90}$ from \cite{2015arXiv150201589P}, we find $\tr=\num[group-digits = false]{2.56069}.$ \\

To describe perturbations about  this model,  we follow the formalism of for example \rfcite{2009PhR...475....1M}, and allow linear perturbations to the  metric.  To linear order, tensor, vector and scalar perturbations decouple.  We therefore can focus solely on tensor perturbations and  the metric then has a line element
\ba 
ds^2=a^2\big(-d\tau^2+\left[\delta_{ij}+h_{ij}\right]dx^i dx^j\big)\,,
\ea
where $h_{ij}$ are transverse, traceless metric tensor perturbations which depend on both $\tau$ and the co--moving spatial co-ordinate $\vec{x}.$ At this linear order, transverse, traceless tensor perturbations do not depend on any choice of gauge.\\

The tensor perturbations have two independent  polarisations,  $+$ and $\times,$
\bas
\label{eq:dof}
h^+_{ij}=h^+e^+_{ij}\qquad\text{and}\qquad
h^\times_{ij}=h^\times e^\times_{ij}\,,
\eas
where the polarisation matrices $e^+_{ij}$ and $e^\times_{ij}$
represent eigenmodes of the spatial Laplacian, each satisfying
\ba
\nabla^2\, e_{ij}= -k^2\,e_{ij}\,,
\ea
and the Laplacian's derivatives are with respect to the co--moving
co--ordinate $\vec{x},$  with $k$ being the perturbation's
co--moving wave--number.\\

The governing equation for the tensor perturbations with neutrino
anisotropic stress is given, in general, by (see e.g.~MalikWands, Rebhan)
\ba
\label{eq:gov-general}
h_{ij}''(\tau)+2\H h_{ij}'(\tau)+\nabla^2 h_{ij}(\tau)=16 \pi G {\pi}^{(\nu)}_{ij}\,,
\ea
where ${\pi}^{(\nu)}_{ij}$ denotes neutrino anisotropic stress, which obeys the constraints
\ba
\label{eq:stress}
\pi^{i}_{i} = 0\,, \qquad\text{and}\qquad \partial_{i}\pi_{ij} = 0\,.
\ea
For a perfect fluid $\pi_{ij} = 0$, but this is not true in general.


\section{Governing equations}
\label{gov_equ_sect}

We can now rewrite the general evolution equation for the tensor
perturbations, \eref{eq:gov-general} above, for the particular cases we
would like to find solutions. We start with simplest case, in which
the anisotropic stress is vanishing.

\subsection{Perfect fluid}
\label{subsec:perfect-fluid}

\subsubsection{Radiation and dust}

As set out in, for example, \rfcite{2009PhR...475....1M}, for a model
where the energy density is in the form of perfect fluids, the tensor
perturbation parameters, $h^+$ and $h^\times,$ from \eref{eq:dof} each
are separately governed by the conformal time equation
\ba\label{eq:gov}
h''(\tau)+2\H h'(\tau)+k^2h(\tau)=0\,,
\ea
where $\H$ is again the conformal Hubble parameter and the dashes
denote differentiation with respect to $\tau.$ Using
\eref{eq:h-in-tau}, we can therefore write \eref{eq:gov} explicitly in
terms of $\tau$ as
\ba
\label{eq:gov-tau}
h''(\tau)+\frac{4\left(2+\tau\right)}{\tau\left(4+\tau\right)}\, h'(\tau)+k^2\, h(\tau)=0\,.
\ea
This is the governing equation for linear tensor perturbations in a
flat universe filled with radiation and pressureless matter, without a
cosmological constant, and with no source term due to anisotropic
stress.

\subsubsection{Radiation only}
\label{subsec:radonly-perfect-fluid}

Similarly, we can derive the governing equation for linear tensor perturbations in flat universes which contain either only radiation or only matter.  Unlike, \eref{eq:gov-tau},  the governing equations in each of these simpler models has  tractable, and well known, analytical solutions.

\begin{wrapfigure}{R}{0.45\linewidth}
\centering
\includegraphics[width=1\linewidth]{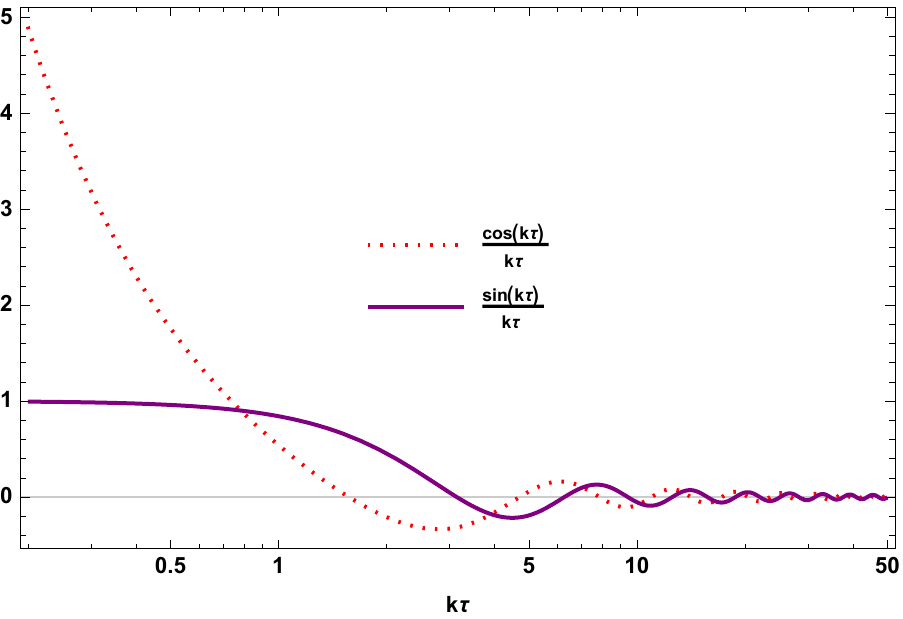}
\caption{Components of the analytical solution for linear tensor perturbations in a flat radiation--only universe, as set out in~\eref{eq:flat-rad-only-soln} for a perfect fluid. The $k\tau$ axis is logarithmic.}
\label{fig:Fig1-sin-cos}
\end{wrapfigure}

For example, with only radiation, the Friedmann equation can be solved
to show that $a\propto \tau.$ We then have
$\H=\infrac{1}{\tau},$ giving us the governing equation for our
perturbation, $\hr,$ as
\ba
\label{eq:gov-tau-rad-only}
\hr''(\tau)+\frac{2}{\tau}\, \hr'(\tau)+k^2\, \hr(\tau)=0\, .
\ea
As recalled in, for example, \rfcite{1993PhRvD..48.4613T,1996PhRvD..53..639W,1994PhRvD..50.3713A,1995PhRvD..52.2112N,2005AnPhy.318....2P,2005astro.ph..5502B}, it is well--known that this can easily be solved analytically to give
$\hr$ as a linear combination
\ba \label{eq:flat-rad-only-soln}
\hr(\tau)=A_\text{s}\,\frac{\sin(k\tau)}{k\tau}+A_\text{c}\,\frac{\cos(k\tau)}{k\tau}\, ,
\ea
where $A_\text{s}$ and $A_\text{c}$ are real constants set by  initial conditions.

This solution also describes the evolution of gravitational waves
which originate very early in the history of a flat radiation--matter
universe, when the matter density can be neglected relative to the
radiation density. Any primordial $A_\text{c}$ component may be
neglected because it decays very rapidly in early times, as can bee
seen from Figure~\ref{fig:Fig1-sin-cos}. In
Section~\ref{sec:modelling-gravitational-waves-through-the-whole-radiation--matter-epoch},
we will use this property to allow us to neglect the $A_\text{c}$
component for gravitational waves which have a very early cosmological
origin, such as inflation.

For use in the following sections, we recall that a perturbation with wave--number $k$ is said to cross the horizon at the time $\tau_k$ when $\H=k.$  From \eref{eq:h-in-tau}  we get 
\ba \label{eq:horizon}
\tau_k=\frac{\sqrt{4 k^2+1}-2 k+1}{k}=\frac{1}{k}+\order\left({\frac{1}{k^2}}\right),
\ea
where the $\order\left({\infrac{1}{k^2}}\right)$ means that the remaining part of the expression is of order $\infrac{1}{k^2}$.

\subsection{Including anisotropic stress}
\label{subsec:including-stress}

\subsubsection{Radiation and dust, including anisotropic stress}

From \ref{subsec:radonly-perfect-fluid} we have an analytical solution
for the tensor perturbations in a perfect fluid. Here we denote how we
can extend this to include neutrino anisotropic stress, we first
recall the expression for the anisotropic stress given in
\rfcite{PhysRevD.50.2541},
\ba \label{eq:stress-int}
\pi_{ij}(\tau) = -4 \bar{\rho}_{\nu}(\tau)\H^{2} \int_{0}^{\tau}K(\tau-t)h'_{ij}(t)dt, 
\ea
where $K$ is the kernel defined as
\ba
\begin{split}
\label{eq:kernel}
K(\tau) &\equiv \frac{1}{16}\int_{-1}^{1}dx(1-x^2)^2 e^{i x \tau}, \\
&= -\frac{\sin{\tau}}{\tau^{3}} -\frac{3\cos{\tau}}{\tau^{4}}+\frac{3\sin{\tau}}{\tau^{5}}.
\end{split}
\ea
and $\bar{\rho}_{\nu}$ is the unperturbed neutrino energy density. To continue we then use \eref{eq:stress-int} in \eref{eq:gov-general}. This gives the integro-differential equation for $h_{ij}(\tau)$ firstly derived in \rfcite{PhysRevD.50.2541} and popularised by \rfcite{2004PhRvD..69b3503W}
\ba
\label{eq:gov-tau-stress}
h''(\tau)+\frac{4\left(2+\tau\right)}{\tau\left(4+\tau\right)}\,h'(\tau)+k^2\, h(\tau)=-24 f_{\nu}(\tau) \left[\frac{4\left(2+\tau\right)}{\tau\left(4+\tau\right)}\right]^2\ \int_{0}^{\tau}K(\tau-t)h'(t)dt.
\ea
This is the governing equation for linear tensor perturbations in a flat  universe filled with radiation and pressureless matter, without a cosmological constant, and including a source term due to anisotropic stress, where $f_{\nu} = \bar{\rho}_{\nu}/\bar{\rho}$. 

\subsubsection{Radiation only with anisotropic stress}
\label{subsec:radonly-stress}

For a radiation only universe \eref{eq:gov-tau-stress} reduces to
\ba
\label{eq:gov-tau-stress-radonly}
\hr''(\tau)+\frac{2}{\tau}\,\hr'(\tau)+k^2\, \hr(\tau)=-24 f_{\nu}(\tau) \left[\frac{4}{\tau^2}\right]\ \int_{0}^{\tau}K(\tau-t)\hr'(t)dt.
\ea
We proceed to use the method from \cite{DPS} to get the solution for the first order $h$ sourced by free--streaming neutrinos to compare with the analytic result \eref{eq:flat-rad-only-soln}. This gives
\ba \label{eq:flat-rad-only-soln-stress}
{\hr}_{(1)}(\tau)= \left( -\frac{i}{k}\right)\left( \frac{e^{i k \tau - f_{\nu}(0)\tau}}{\tau}\right),
\ea
which is less than $1\%$ away from the analytic solution without anisotropic stress.

\section{Analytical solutions for the  tensor perturbations}
\label{sec:approximating-tensor-perturbations}

  In this section, we will derive an approximate analytical solution to the governing equation \eref{eq:gov-tau} valid in  the \emph{sub--horizon} case, $k\gg\H.$  We begin by recalling the method of \rfcite{FSN} as applicable to a single second order differential equation, having a small parameter. Here, the small parameter is the co--moving inverse wave--number, $k^{-1}.$

As described in \rfcite{DPS}, the method consists of finding an approximation for $h$ of the form
\ba \label{eq:FSN-approx}
h(\tau)= \exp\left[{\sum_{s=0}^\infty \int k^{-s+1}\,\omega_s(\tau)\,d\tau}\right],
\ea
  and for  convenience, we write
\ba\label{eq:omega}
\omega&=\sum_{s=0}^\infty k^{-s+1}\,\omega_s(\tau),
\ea
where $\omega$ depends on both $\tau$ and $k,$ and we will regard $k$ as a fixed parameter. We then have
\bas
h'&=\omega h,\\
\mathrm{and\quad\quad} h''&=\left(\omega'+\omega^2\right)h\,  . 
\eas

\subsection{Approximating perfect fluid tensor perturbations}
\label{sec:approximating-tensor-perturbations}
\begin{wrapfigure}{R}{0.45\linewidth}
\centering
\includegraphics[width=1\linewidth]{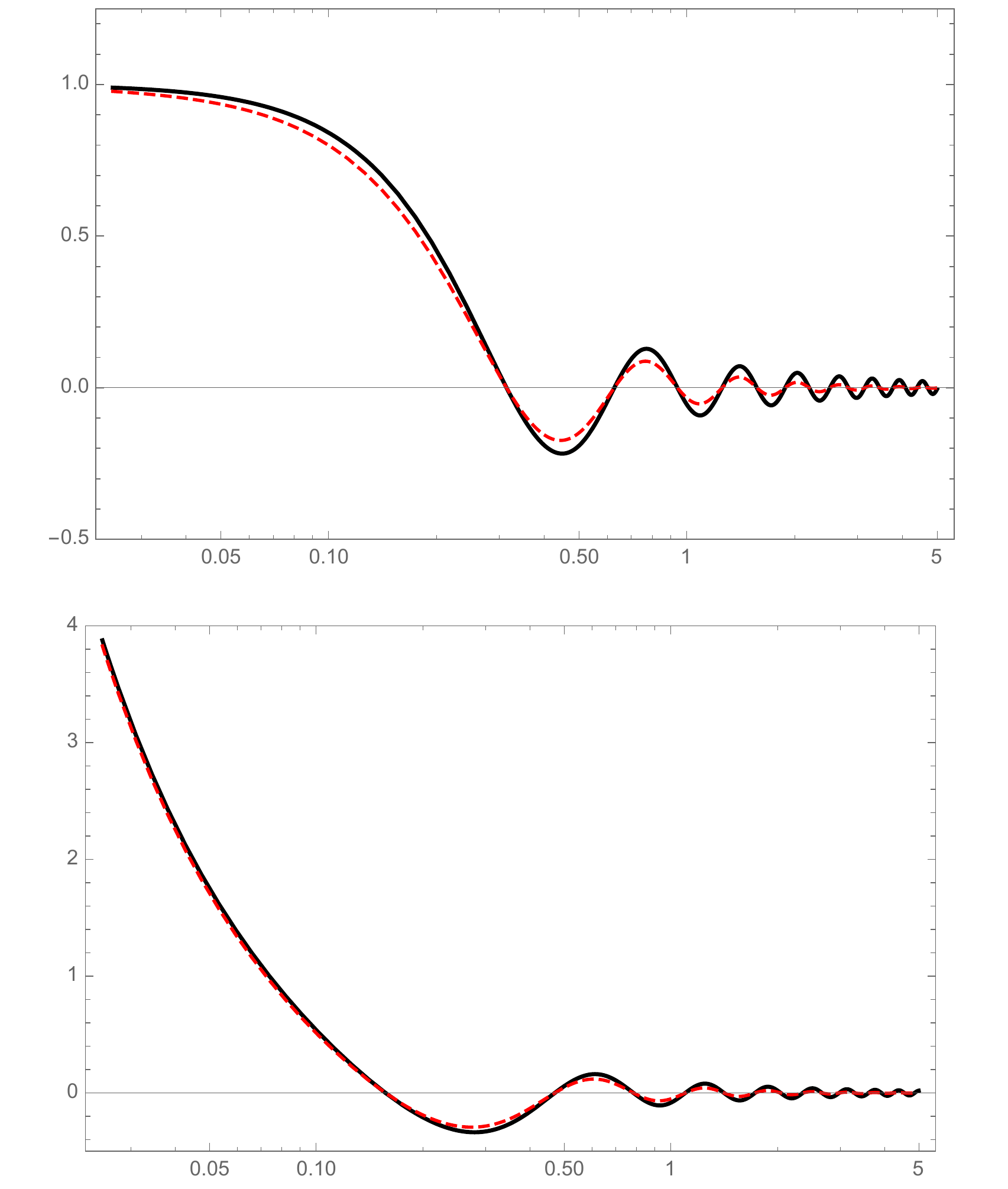}
\caption{\emph{Upper panel:} Real components of the analytical and approximate solution for linear tensor perturbations in a flat radiation--only universe without (solid black line red dashed line) and with the effect of free streaming neutrinos anisotropic stress with $f_{\nu}(0)=0.5$  (red dashed line), as set out in Eqs.~(\ref{eq:flat-rad-only-soln}) and (\ref{eq:flat-rad-only-soln-stress}). \emph{Lower panel:} Imaginary components of Eqs.~(\ref{eq:flat-rad-only-soln}) and (\ref{eq:flat-rad-only-soln-stress}). The $k\tau$ axis is logarithmic.}
\label{fig:Fig2-sin-cos-stress}
\end{wrapfigure}

Using the approximation of \eref{eq:FSN-approx} in our governing equation \eref{eq:gov-tau}, and dividing through by $h,$ we get  the ``characteristic'' equation from which we will derive all orders of our approximation, 
\ba \label{eq:approxequation}
\left(\omega'+\omega^2\right)+\frac{4\left(2+\tau\right)}{\tau\left(4+\tau\right)}\,\omega+k^2=0\, .
\ea
We  can now  substitute \eref{eq:omega} into  \eref{eq:approxequation} and equate coefficients of like powers of the inverse wave--number, $k^{-1}$.  The calculation is similar to that set out in a Bessel type differential equation in \rfcite{DPS}. It can also be carried out using the \emph{Mathematica} notebook available online at \rfcite{GWGithub}.

We can form successive order $\smax$ approximations 
\ba \label{eq:smax}
h_{(\smax)}
=\exp\left[\sum_{s=0}^{\smax}
\int k^{-s+1}\,\omega_s(\tau)\,d\tau\right].
\ea
We find that $h_{(\smax)}$ has complex values.   Taking the real and imaginary parts of $h_{(\smax)}$ provides approximations for two independent solutions of \eref{eq:gov-tau}.

For clarity, we should point out that, in this paper, when we describe an approximation as being of a particular order, we are not referring to the order of the perturbation theory (as described in, for example, \rfcite{2009PhR...475....1M}). We use the term order in this context to refer to the value of $\smax$ in the approximation of \eref{eq:smax}. We only use expressions of first order in perturbation theory, and, to minimise confusion, we consistently describe these as being of linear order.

The largest value of $\smax$ we shall use is $\smax=3.$  This is because  trial against numerical solutions\footnote{  In Section~VII of \rfcite{DPS} we set out a method for estimating which order approximation is the most accurate for a given value of $\tau$ without employing numerical solutions. } shows that  larger choices of $\smax$ do not uniformly improve the approximation for all sub--horizon values of $\tau.$ 

This third order approximation is to take any linear combination of the real and imaginary parts of
\ba \label{eq:third-order}
h_{(3)}(\tau)=\frac{1}{\tau(4+\tau )}\left(\frac{4+\tau }{\tau }\right)^{\infrac{i}{4 k}}\exp\left[i k \tau+\frac{1}{2k^2\tau \left( 4+\tau  \right)}\right].
\ea
Recall, as set out following \eref{eq:rmscalefactor}, that $\tau$ and $k$ have been normalised in terms of $\tunit$ and are therefore dimensionless numbers.

We want to compare our approximations with numerical solutions.\footnote{As usual, our numerical solutions are derived from initial conditions.  In \rfcite{DPS} we had to use final conditions to manage a particular numerical instability.} 
Figure~\ref{fig:Fighappr3k10} shows the $h_{(3)}$ approximation for wave--number $k=10\kunit,$ plotted against the corresponding numerical solution.

  We want now to quantify the error in the approximation $h_{(3)}.$ One way to try to do this would be to take the numerical solution $h$ and calculate the ratio $\infrac{\left|\left(h_{(3)}-h\right)}{h}\right|.$  However, this runs into a difficulty.  Since $h$ is oscillating and repeatedly takes zero values, unless there is no error at all at these zeros, the ratio $\infrac{\left|\left(h_{(3)}-h\right)}{h}\right|$ will repeatedly become infinite.  

 We therefore adapt our approach in order to avoid this problem.  Broadly speaking, we estimate the typical error compared with the amplitude of oscillation.  To do this we first decide a target degree of accuracy, $1\%$ say.  We then compare a plot of the error $\left|h_{(3)}-h\right|$ with a plot of the target accuracy, here $1\%\cdot \left|h\right|.$  Both plots will usually spike repeatedly downwards towards zero.  We consider that the approximation is accurate to within $1\%$ if, spikes when $h=0$ aside, $\left|h_{(3)}-h\right|\le 1\%\cdot \left|h\right|.$  The will show on the graph as the plot of $\left|h_{(3)}-h\right|$ being level with, or below, the plot of $1\%\cdot \left|h\right|$ (except near $h=0$ spikes). 

As shown in Figure~\ref{fig:Fighappr3k10error},  the $h_{(3)}$  approximation is accurate to within $1\%,$ or better from  horizon--crossing onwards $(\tau\ge\tau_k).$  In fact, this degree of accuracy applies for wave--numbers $k\gtrsim 9\kunit\sim\SI{0.07}{a_0.Mpc^{-1}}.$ Using \eref{eq:actual}, this corresponds to actual present day distances of {$\SI{17}{Mpc}.$ 

\begin{figure}
\centering
\includegraphics[width=1\linewidth]{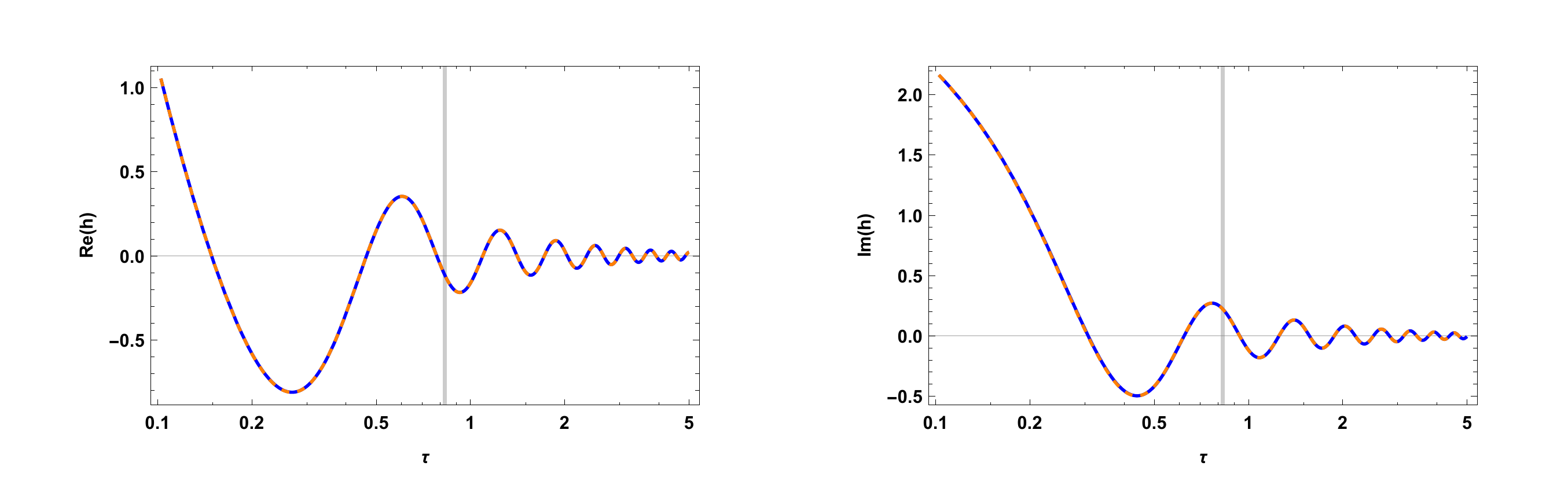}
\caption{  The real and imaginary parts of the approximate third order solution for a perfect fluid given by~\eref{eq:third-order} (solid blue line) and the numerical solution (dashed orange line) for \eref{eq:gov-tau} with $k=10\kunit$. The horizontal axes show $\tau$ in units of $\tau_\text{r},$ the conformal time of recombination, calculated using \rfcite{2015arXiv150201589P}. The curves are plotted for values of $\tau$ after horizon--crossing. The solid vertical line corresponds to radiation--matter equality, as defined by \eref{eq:rmscale}. }
\label{fig:Fighappr3k10}
\end{figure}

\begin{figure}
\centering
\includegraphics[width=1\linewidth]{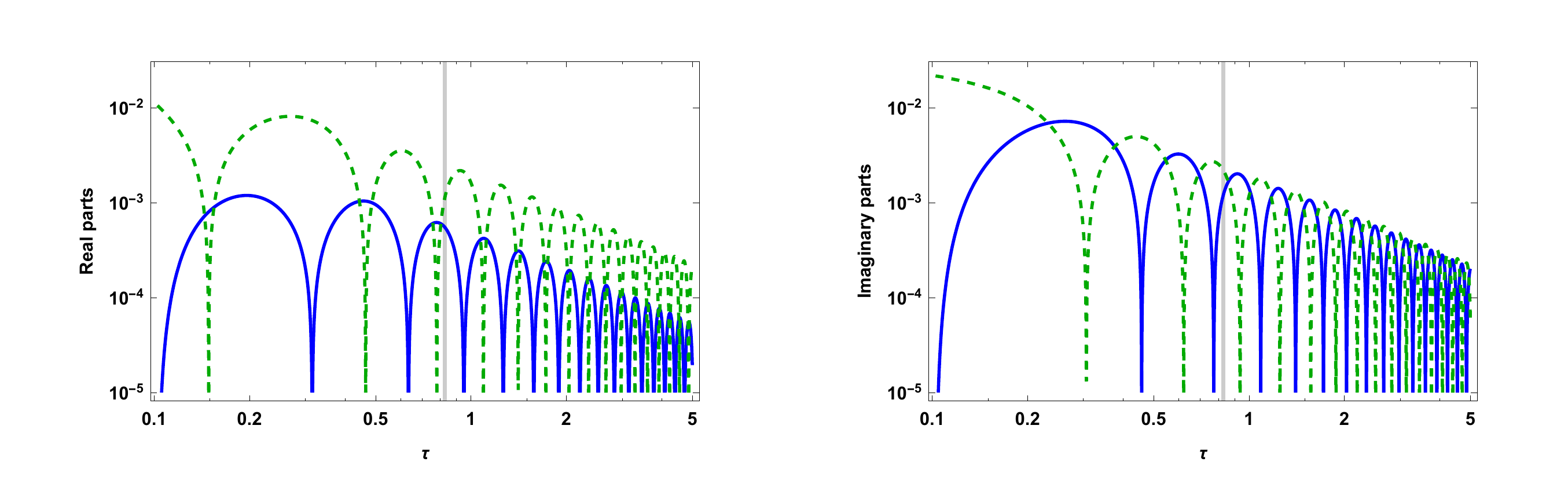}
\caption{  The solid lines show the absolute errors in the approximation of Figure~\ref{fig:Fighappr3k10} as $\left|f_{(3)}(\tau)-f(\tau)\right|,$ where $f_{(3)}$ is the real or imaginary part, as appropriate, of $h_{(3)}$ from \eref{eq:third-order} and $f$ is the real or imaginary relevant part of the numerical solution for a perfect fluid.  For comparison, the dashed line shows 
$\left|1\%\times f(\tau)\right|.$ Downward spikes in the lines represent points where the value becomes zero. The blue solid line being below or level with the green dashed line indicates an error of less than or equal to $1\%.$  See the text for further explanation. }
\label{fig:Fighappr3k10error}
\end{figure}

  The second order and first order approximations constructed by our method will also be of use in the next section, and we set them out here.   The second order approximation is to take any linear combinations of the real and imaginary parts of
\ba \label{eq:second-order}
h_{(2)}(\tau)=\frac{1}{\tau(4+\tau )}\left(\frac{4+\tau }{\tau }\right)^{\infrac{i}{4 k}}\exp\left[i k \tau\right].
\ea
For use in Section~\ref{sec:modelling-gravitational-waves-through-the-whole-radiation--matter-epoch}, we also write this more explicitly as 
\ba \label{eq:second-order-trig}
h_{(2)}(\tau)=\frac{B_\text{s}\sin \Big(k \tau +\frac{1}{4k}\ln \left(1+\frac{4}{\tau }\right)\Big)+B_\text{c}\cos \Big(k \tau +\frac{1}{4k}\ln \left(1+\frac{4}{\tau }\right)\Big)}{\tau  (\tau +4)},
\ea
with $B_\text{s}$ and $B_\text{c}$ real constants. This second order approximation is accurate to within $1\%,$ or better, from horizon-crossing onwards,  when we have $k\gtrsim 17\kunit\sim\SI{0.12}{a_0.Mpc^{-1}}.$

The first order, or leading order, approximation~--- as set out in \cite{2008cosm.book.....W}~--- is to take any linear combination of the real and imaginary parts of
\ba \label{eq:leading}
h_{(1)}(\tau)=\frac{1}{\tau(4+\tau )}\exp\left[i k \tau\right]
\propto
\frac{\exp\left[i k \tau\right]}{a(\tau)}
=\frac{\cos\left( k \tau\right)+i\sin\left( k \tau\right)}{a(\tau)},
\ea
  which  is accurate to $1\%,$ or better, from horizon-crossing onwards, when we have $k\gtrsim 120\kunit\sim\SI{0.87}{a_0.Mpc^{-1}}.$

\subsection{Approximating tensor perturbations with anisotropic stress}
\label{sec:approximating-tensor-perturbations-with-stress}

In this section, we compute an approximate analytical solution to the governing equation \eref{eq:gov-tau-stress}. The fraction of the total energy density in neutrinos is \cite{2004PhRvD..69b3503W} 
\ba
\label{eq:f-neutrinos}
f_{\nu}(\tau) =
\frac{\Omega(\infrac{a_{0}}{a})^{4}}{\Omega_{\rm{m}}(\infrac{a_{0}}{a})^{3}+\left(\Omega_{\gamma} + \Omega_{\nu}\right) (\infrac{a_{0}}{a})^{4}} = \frac{f_{\nu}(0)}{1+\tau},
\ea

We follow the method described in \rfcite{DPS} along with (\ref{eq:kernel}) and (\ref{eq:f-neutrinos}) to get an approximate solution to \eref{eq:gov-tau-stress}. The third order approximation is to take any linear combination of the real and imaginary parts of 
\ba 
\label{eq:third-order-stress}
\begin{split}
h_{(3)}(\tau)&= \left( -\frac{i}{k} - \frac{2(2+\tau)}{k^{2} \tau (4+\tau)} + \frac{i[16 f_{\nu}(0)(2+\tau)^{2} + 5(16+28\tau+15\tau^{2}+3\tau^{3})]}{5 k^{3} \tau^{2}(1+\tau)(4+\tau)^{2}} \right) \\
&\qquad \times \tau^{-(1 + (i/k)\alpha)} (1+\tau)^{(i/k) \beta} (4+\tau)^{-1+(i/k) \gamma} \\
&\qquad \times \exp\left[ i k \tau - f_{\nu}(0)\tau- \frac{2 i [-15(2+\tau)+4 f_{\nu}(0)(6+\tau)]}{15 k \tau (4+\tau)}+\frac{16 f_{\nu}(0)(2+\tau)^{2} + 5(16+28\tau+15\tau^{2}+3\tau^{3})]}{10k^{2} \tau^{2}(1+\tau)(4+\tau)^{2}}\right] .
\end{split}
\ea
where
\ba
\alpha = \frac{5+8 f_{\nu}(0)}{20}, \quad \beta = \frac{16}{45}f_{\nu}(0), \quad \text{and} \quad \gamma = \frac{45+8 f_{\nu}(0)}{180} .
\ea
\begin{figure}
\centering
\includegraphics[width=1\linewidth]{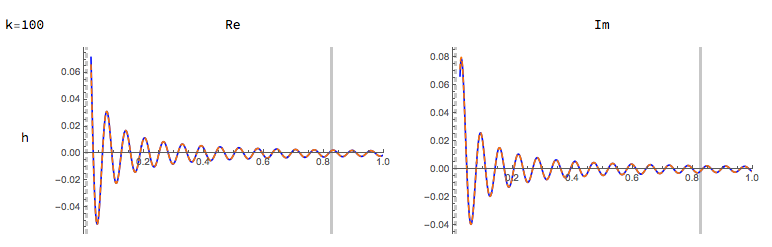}
\caption{The real and imaginary parts of the approximate third order solution with free neutrino anisotropic stress given by~\eref{eq:third-order-stress} (solid blue line) and the numerical solution (dashed orange line) for \eref{eq:gov-tau-stress} with $k=100\kunit$ and $f_{\nu}(0)=0.40523$. The curves are plotted for values of $\tau$ after horizon--crossing. The solid vertical line corresponds to radiation--matter equality, as defined by \eref{eq:rmscale}.}
\label{fig:Fighappr5k100}
\end{figure}

\section{Modelling primordial gravitational waves}
\label{sec:modelling-gravitational-waves-through-the-whole-radiation--matter-epoch}

In this section, we model the propagation of primordial gravitational
waves from early times when the waves are super--horizon, through to
later times when they are sub--horizon.  We do this by matching an
approximate solution for early times with a different one for late
times.  This gives us a solution valid for all times in the
radiation--matter model, not just when the waves are sub--horizon.

\begin{wrapfigure}{R}{0.45\linewidth}
\centering
\includegraphics[width=1\linewidth]{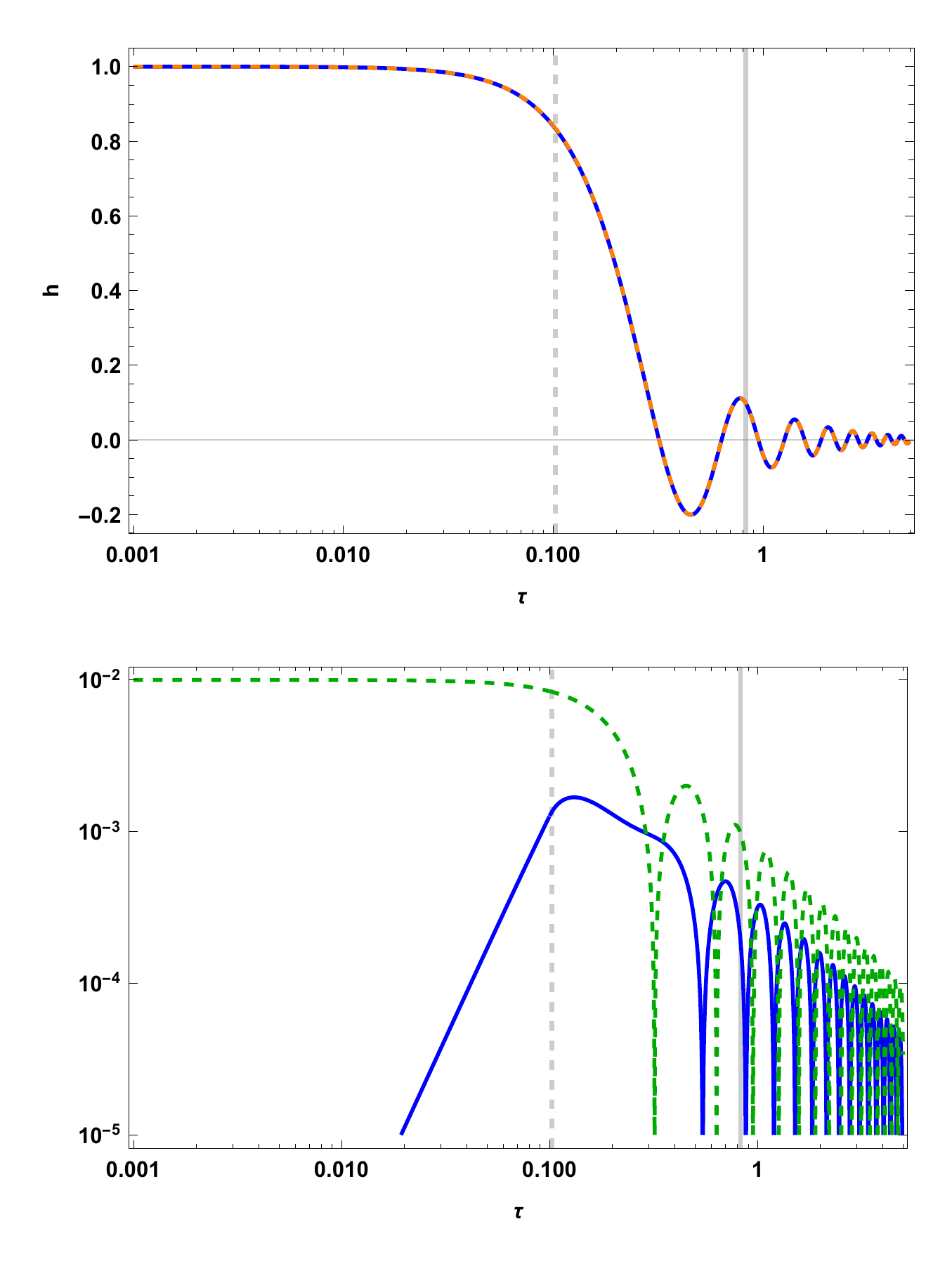}
\caption{\emph{Upper panel:} The approximate solution, $\hmatch,$ of
  ~\eref{eq:match} (solid blue line) and the numerical solution for a perfect fluid
  (dashed orange line) for \eref{eq:gov-tau} with $k=10\kunit$.
  \emph{Lower panel:} The error in the matched solution (solid blue
  line) plotted against $1\%$ of the numerical solution (dashed green
  line). \emph{In both panels:} The vertical dashed line is
  horizon--crossing, $\tau_{k=10}\approx \num{0.102},$ which is very
  close to the matching point, $\tau=\num{0.1}.$ The solid vertical
  line is radiation--matter equality.  }
\label{fig:Fig-match}
\end{wrapfigure}

For early times, we use the radiation--only solution,
\eref{eq:flat-rad-only-soln}, while for late times we use the previous
section's approximate solution $h_{(2)}$ from
\eref{eq:second-order-trig}.  We choose the time at which we match
these solutions to be $\tau=\infrac{1}{k}.$ We note from
\eref{eq:horizon} that for relevant values of $k,$ the matching time
$\tau=\infrac{1}{k}$ will be close to the horizon--crossing time
$\tau_k.$ The form of the resulting equations will, however, be much
simpler than if we had done the matching at the horizon--crossing time
itself.

 The derivation of the approximations set out in the previous section depended on the sub--horizon assumption, $k\ll\H.$  However, below we show that, for $k\gtrsim 4.5\kunit,$  our matching approach gives an approximation, $\hmatch,$ which is accurate to $1\%,$ or better, through the whole radiation--matter epoch.  We use the second order solution $h_{(2)}$ in the matching because trial and error shows it is the order of approximate solution which works to within $1\%$ for the widest range of wave--numbers. 

From~\eref{eq:flat-rad-only-soln}, we have that the early time solution is 
\ba \label{eq:flat-rad-only-soln2}
\hr(\tau)=A_\text{s}\,\frac{\sin(k\tau)}{k\tau}+A_\text{c}\,\frac{\cos(k\tau)}{k\tau}.
\ea
As discussed above following \eref{eq:flat-rad-only-soln}, for gravitational
waves with primordial origin, we can set $A_\text{c}=0.$ We therefore
take
\ba \label{eq:flat-rad-only-soln3}
\hr(\tau)=\frac{\sin(k\tau)}{k\tau}
\ea
to be our solution for $\tau\le\infrac{1}{k}$.

 From \eref{eq:second-order-trig}, we have our solution for $\tau\ge\infrac{1}{k}$  as
\ba \label{eq:second-order-trig2}
h_{(2)}(\tau)=\frac{B_\text{s}\sin \left(\lambda(k,\tau)\right)+B_\text{c}\cos \left(\lambda(k,\tau)\right)}{\tau  (\tau +4)},
\ea
where we have written 
\be \label{eq:lamdba}
\lambda(k,\tau)=k\tau+\frac{1}{4 k}\ln \left(1+\frac{4}{\tau}\right).
\ee

 We then need to choose the real constants $B_\text{s}$ and $B_\text{c}$ in \eref{eq:second-order-trig2}  to be such that $\hmatch$ and $\hmatch'$  are continuous. In other words, we ensure the values of the two functions and the values of their first $\tau$ derivatives each match at $\tau=\infrac{1}{k}.$ The calculation is set out in the \hyperref[sec:calculation-of-the-matching-conditions-for-approximating-primordial-gravitational-waves]{Appendix}.  From \eref{eq:match-a}, we have the resulting matching approximation 
\begin{equation}\label{eq:match}
\hmatch(\tau)=
\begin{dcases}
\frac{\sin(k\tau)}{k\tau}
&\mbox{if } \tau\le\frac{1}{k}
\\[9pt]
\frac{4 k+1}{4 k^3 \tau  (4+\tau)}
\bigg[
\,4 k\, \sin \Big(k\tau + L(k,t)\Big)
+ \mu
\sin \Big(k\tau + L(k,\tau)-1\Big)
\bigg]
 &\mbox{if } \tau\ge\frac{1}{k},
\end{dcases}
\end{equation}
where 
\be
L(k,\tau)=\frac{1}{4 k}\ln \left(\frac{1+4\,\tau^{-1} }{1+4\,k\hfill}\right) 
\qquad
\text{
and }\qquad
\mu
=
\sin\left(1\right)+\cos\left(1\right)
=
1.38177...\, .
\ee

 This is our approximate analytical solution to \eref{eq:gov-tau}, the governing equation for tensor perturbations (gravitational waves) in a flat radiation--matter universe.  As we now show, it is a good approximation both sub--horizon ($\tau\lesssim\infrac{1}{k}$) and super--horizon ($\tau\gtrsim\infrac{1}{k}$) for wave--numbers $k$ over a wide range of values relevant for structure formation in the Universe.

Figure~\ref{fig:Fig-match} shows this matched solution, $\hmatch,$ for
$k=10\kunit$ in its upper panel and in the lower panel plots the error
using the approach also used in Figure~\ref{fig:Fighappr3k10error}.
The lower panel shows that the error is consistently less than $1\%,$
both before and after horizon--crossing.  We found that $\hmatch$ has
errors of $1\%,$ or better for all $\tau$, when we $k\gtrsim
4.5\kunit=\SI{0.033}{a_0.Mpc^{-1}}.$

Using \eref{eq:actual},   this corresponds to $\hmatch$ meeting   our $1\%$ accuracy test for all times, when we consider inverse wave--numbers with actual present day values of less than, or equal to, around $\SI{35}{Mpc}.$  We can also express this in multipoles, as used in CMB calculations, via the broad correspondence ~--- see for example \rfcite{2003itc..book.....R,2008cosm.book.....W,2009pdp..book.....L} 
~---   that a multipole, $l,$ receives its main contributions from present day physical wave--numbers $k_\text{ph}\simeq {c H_0} \,l.$ Using this rule, our range of $1\%$ accuracy, $k\gtrsim 4.5\kunit,$ broadly corresponds to multipoles $l\gtrsim 120.$

We found that matching using $h_{(1)}$ of \eref{eq:leading}, in place of $h_{(2)},$ obtains errors of $1\%,$ or better for all $\tau$, only when we have  $k\gtrsim 180\kunit=\SI{1.3}{a_0.Mpc^{-1}}.$ That corresponds to actual present day inverse wave--numbers of less, or equal to, around than $\SI{0.9}{Mpc},$ and, broadly, to multipoles $l\gtrsim\num{5000}.$\\ 


 In the \hyperref[sec:intro]{Introduction}, we referred to  the WKB solutions of \rfcite{1995PhRvD..52.2112N,2005AnPhy.318....2P}, \rfcite{2005astro.ph..5502B} and \rfcite{2008cosm.book.....W}.  These aim to derive approximations for primordial gravitational waves. We now compare these with our approximation~--- not restricting our attention only to the perfect fluid case but considering the anisotropic stress due to free--streaming neutrinos as well like \rfcite{2005AnPhy.318....2P,2008cosm.book.....W,2005astro.ph..5502B} .

 Turning first to the WKB approximation of  \rfcite{2008cosm.book.....W}, we note that this is the same as our leading order approximation $h_{(1)}$ of  \eref{eq:leading}.  For primordial gravitational waves, along the lines discussed following \eref{eq:flat-rad-only-soln}, we are interested in only the imaginary, sine, part of \eref{eq:leading}, which is proportional to
\ba \label{eq:leading-s}
h_{(1),\text{s}}=\frac{\arm \sin(k\tau)}{k\, a(\tau)}=\frac{4\sin(k\tau)}{k\tau(4+\tau)},
\ea
where we have chosen the constant of proportionality so that $h_{(1),\text{s}}(0)=1.$
Figure~\ref{fig:Fig-match-happr1} shows that, for $k=10\kunit,$ this gives errors of around $10\%.$   

\begin{wrapfigure}{R}{0.45\linewidth}
\centering
\includegraphics[width=1\linewidth]{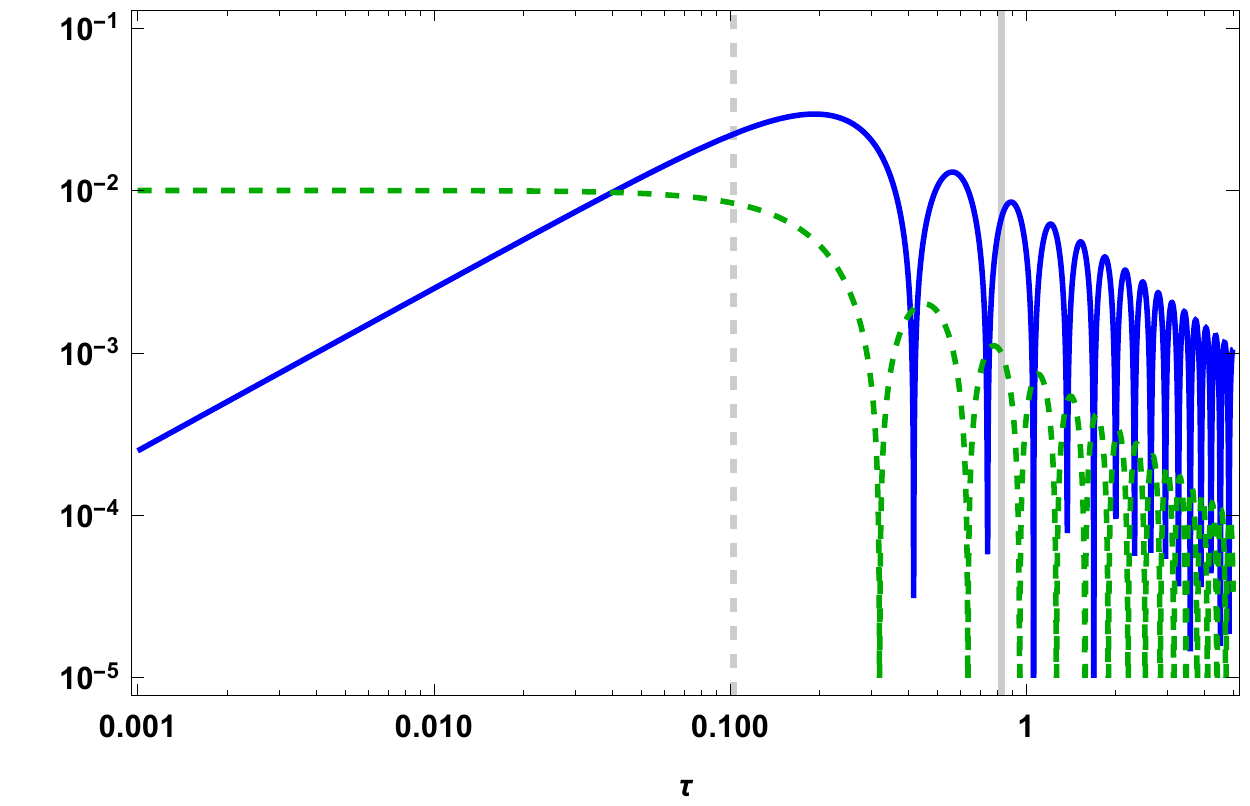}
\caption{As for the lower panel of Figure~\ref{fig:Fig-match}, but using the leading order approximation, $h_{(1),\text{s}}$ of \eref{eq:leading-s}, instead of $\hmatch.$}
\label{fig:Fig-match-happr1}
\end{wrapfigure}

References~\cite{1995PhRvD..52.2112N,2005AnPhy.318....2P} presented
another less straightforward form of WKB approximation.  Figures in
those papers show that, for $k\approx 16.4\kunit$ in Fig.~1 of
\rfcite{1995PhRvD..52.2112N}, and for $k\approx 12.1\kunit$ in Fig.~2
of \rfcite{2005AnPhy.318....2P}, there are errors of around $10\%,$ or
greater.  Their figures also show that alternative simpler
approximations~--- using a radiation--only solution, a matter--only
solution, or the two matched together at $\tau=\tau_\text{eq}$ in a
``sudden approximation''~--- are all less accurate than the WKB
approximation of \rfcite{1995PhRvD..52.2112N,2005AnPhy.318....2P}, and
so also much less accurate than our approximation $\hmatch$, given in
\eref{eq:match}.

Our calculations suggest that the best approximation we have found in the literature for sub--horizon primordial gravitational waves is the WKB solution explored in passing in \rfcite{2005astro.ph..5502B}. That approximation is constructed for only one of the two independent solutions of the governing equation \eref{eq:gov-tau}.  As can be seen from Figure~\ref{fig:Fighappr3k10error}, typically one of a pair of approximate WKB solutions will be better than the other.  For $k=10\kunit,$ and for sub--horizon values of $\tau,$ the approximation of \rfcite{2005astro.ph..5502B} is accurate to within about $0.8\%,$ which compares with the accuracies for the two approximations of Figure~\ref{fig:Fighappr3k10error} of around $0.25\%$ (real part) and $1\%$ (imaginary part).  When started outside the horizon, \rfcite{2005astro.ph..5502B}'s approximation becomes considerably less accurate than our matching approximation, $\hmatch.$ 

The Bessel function approach of \rfcite{2013PhRvD..88h3536S} also approximates primordial gravitational waves.  As indicated in  Section~\ref{sec:intro}, a comparison with numerical solutions shows that many terms are needed to derive  accurate  approximations.

\section{Conclusion}
\label{sec:conclusion}

In this work we have derived new analytic solutions to the
gravitational wave or tensor evolution equation at linear order in
cosmological perturbation theory, using the method presented in
\rfcite{DPS} and including neutrino anisotropic stress. The solutions
depend on time, wavenumber, and the ratio of the neutrino background
energy density and the total energy density at $\tau=0$, $f_{\nu}(0)$,
which allows to vary the damping of the gravitational waves due to the
neutrino anisotropic stress.

We show how the method of \rfcite{DPS} can be applied to a
integro-differential equation, and our solutions present a simple way
to analyse the evolution of the tensor perturbations in the presence
of anisotropic stress. The analytic solutions save computational time
compared to solving the equations numerically, for example if we are
interested in a wide range of different parameter values $f_\nu$, the ratio of neutrino to total background energy density.
%
We have compared our analytical approximation to numerical solutions
with anisotropic stress and find that the difference between them is
within $1 \%$.


A simple example will illustrate the usefulness of analytical
solutions even in this day and age.  In second order cosmological
perturbation theory, the governing equations include terms of the form
$h_{ij}h^{ij}$.
For example it has been shown in \rfcite{2016JCAP...02..021C} that
different forms of the curvature perturbation at second order differ
by terms proportional to $h_{ij}h^{ij}$. To relate this to the
expressions derived in the sections above, we can use that for a
gravitational wave travelling in the $z$--direction, for example,
\bea
h_{ij}h^{ij} = 2\left[\left(h^{\times}\right)^2+\left(h^{+}\right)^2\right]\,,
\eea
where $h^{\times}$ and $h^{+}$ are the two independent polarisations
of the gravitational waves defined in \eref{eq:dof}. Then substituting
in the solutions of the model we are interested in, we get an
expression of how much the different curvature perturbations disagree
in terms of, say, the wavenumber and the neutrino-radiation ratio
$f_\nu$.  We shall return to these issues in future work.

\section*{Acknowledgements}

%
The authors are grateful to Pedro Carrilho for useful discussions. 
KAM is supported, in part, by STFC grant ST/J001546/1.
JLF acknowledges support of a studentship funded by Queen Mary
University of London as well as CONACYT grant No.~603085.



\bibliographystyle{apsrev4-1}

\begin{thebibliography}{22}%
\makeatletter
\providecommand \@ifxundefined [1]{%
 \@ifx{#1\undefined}
}%
\providecommand \@ifnum [1]{%
 \ifnum #1\expandafter \@firstoftwo
 \else \expandafter \@secondoftwo
 \fi
}%
\providecommand \@ifx [1]{%
 \ifx #1\expandafter \@firstoftwo
 \else \expandafter \@secondoftwo
 \fi
}%
\providecommand \natexlab [1]{#1}%
\providecommand \enquote  [1]{``#1''}%
\providecommand \bibnamefont  [1]{#1}%
\providecommand \bibfnamefont [1]{#1}%
\providecommand \citenamefont [1]{#1}%
\providecommand \href@noop [0]{\@secondoftwo}%
\providecommand \href [0]{\begingroup \@sanitize@url \@href}%
\providecommand \@href[1]{\@@startlink{#1}\@@href}%
\providecommand \@@href[1]{\endgroup#1\@@endlink}%
\providecommand \@sanitize@url [0]{\catcode `\\12\catcode `\$12\catcode
  `\&12\catcode `\#12\catcode `\^12\catcode `\_12\catcode `\%12\relax}%
\providecommand \@@startlink[1]{}%
\providecommand \@@endlink[0]{}%
\providecommand \url  [0]{\begingroup\@sanitize@url \@url }%
\providecommand \@url [1]{\endgroup\@href {#1}{\urlprefix }}%
\providecommand \urlprefix  [0]{URL }%
\providecommand \Eprint [0]{\href }%
\providecommand \doibase [0]{http://dx.doi.org/}%
\providecommand \selectlanguage [0]{\@gobble}%
\providecommand \bibinfo  [0]{\@secondoftwo}%
\providecommand \bibfield  [0]{\@secondoftwo}%
\providecommand \translation [1]{[#1]}%
\providecommand \BibitemOpen [0]{}%
\providecommand \bibitemStop [0]{}%
\providecommand \bibitemNoStop [0]{.\EOS\space}%
\providecommand \EOS [0]{\spacefactor3000\relax}%
\providecommand \BibitemShut  [1]{\csname bibitem#1\endcsname}%
\let\auto@bib@innerbib\@empty
\bibitem [{\citenamefont {{Abbott}}\ \emph {et~al.}(2016)\citenamefont
  {{Abbott}}, \citenamefont {{Abbott}}, \citenamefont {{Abbott}}, \citenamefont
  {{Abernathy}}, \citenamefont {{Acernese}}, \citenamefont {{Ackley}},
  \citenamefont {{Adams}}, \citenamefont {{Adams}}, \citenamefont {{Addesso}},
  \citenamefont {{Adhikari}},\ and\ \citenamefont
  {et~al.}}]{2016PhRvL.116f1102A}%
  \BibitemOpen
  \bibfield  {author} {\bibinfo {author} {\bibfnamefont {B.~P.}\ \bibnamefont
  {{Abbott}}}, \bibinfo {author} {\bibfnamefont {R.}~\bibnamefont {{Abbott}}},
  \bibinfo {author} {\bibfnamefont {T.~D.}\ \bibnamefont {{Abbott}}}, \bibinfo
  {author} {\bibfnamefont {M.~R.}\ \bibnamefont {{Abernathy}}}, \bibinfo
  {author} {\bibfnamefont {F.}~\bibnamefont {{Acernese}}}, \bibinfo {author}
  {\bibfnamefont {K.}~\bibnamefont {{Ackley}}}, \bibinfo {author}
  {\bibfnamefont {C.}~\bibnamefont {{Adams}}}, \bibinfo {author} {\bibfnamefont
  {T.}~\bibnamefont {{Adams}}}, \bibinfo {author} {\bibfnamefont
  {P.}~\bibnamefont {{Addesso}}}, \bibinfo {author} {\bibfnamefont {R.~X.}\
  \bibnamefont {{Adhikari}}}, \ and\ \bibinfo {author} {\bibnamefont
  {et~al.}},\ }\href {\doibase 10.1103/PhysRevLett.116.061102} {\bibfield
  {journal} {\bibinfo  {journal} {Physical Review Letters}\ }\textbf {\bibinfo
  {volume} {116}},\ \bibinfo {eid} {061102} (\bibinfo {year} {2016})},\ \Eprint
  {http://arxiv.org/abs/1602.03837} {arXiv:1602.03837 [gr-qc]} \BibitemShut
  {NoStop}%
\bibitem [{\citenamefont {Ade}\ \emph {et~al.}(2015)\citenamefont {Ade},
  \citenamefont {Aghanim}, \citenamefont {Ahmed}, \citenamefont {Aikin},
  \citenamefont {Alexander}, \citenamefont {Arnaud}, \citenamefont {Aumont},
  \citenamefont {Baccigalupi}, \citenamefont {Banday}, \citenamefont
  {Barkats},\ and\ \citenamefont {et~al.}}]{PhysRevLett.114.101301}%
  \BibitemOpen
  \bibfield  {author} {\bibinfo {author} {\bibfnamefont {P.~A.~R.}\
  \bibnamefont {Ade}}, \bibinfo {author} {\bibfnamefont {N.}~\bibnamefont
  {Aghanim}}, \bibinfo {author} {\bibfnamefont {Z.}~\bibnamefont {Ahmed}},
  \bibinfo {author} {\bibfnamefont {R.~W.}\ \bibnamefont {Aikin}}, \bibinfo
  {author} {\bibfnamefont {K.~D.}\ \bibnamefont {Alexander}}, \bibinfo {author}
  {\bibfnamefont {M.}~\bibnamefont {Arnaud}}, \bibinfo {author} {\bibfnamefont
  {J.}~\bibnamefont {Aumont}}, \bibinfo {author} {\bibfnamefont
  {C.}~\bibnamefont {Baccigalupi}}, \bibinfo {author} {\bibfnamefont {A.~J.}\
  \bibnamefont {Banday}}, \bibinfo {author} {\bibfnamefont {D.}~\bibnamefont
  {Barkats}}, \ and\ \bibinfo {author} {\bibnamefont {et~al.}} (\bibinfo
  {collaboration} {(BICEP2/Keck and Planck Collaborations)}),\ }\href {\doibase
  10.1103/PhysRevLett.114.101301} {\bibfield  {journal} {\bibinfo  {journal}
  {Phys. Rev. Lett.}\ }\textbf {\bibinfo {volume} {114}},\ \bibinfo {pages}
  {101301} (\bibinfo {year} {2015})},\ \Eprint
  {http://arxiv.org/abs/1502.00612} {arXiv:1502.00612 [astro-ph.CO]}
  \BibitemShut {NoStop}%
\bibitem [{\citenamefont {{Planck Collaboration}}\ \emph
  {et~al.}(2015)\citenamefont {{Planck Collaboration}}, \citenamefont {{Ade}},
  \citenamefont {{Aghanim}}, \citenamefont {{Arnaud}}, \citenamefont
  {{Ashdown}}, \citenamefont {{Aumont}}, \citenamefont {{Baccigalupi}},
  \citenamefont {{Banday}}, \citenamefont {{Barreiro}}, \citenamefont
  {{Bartlett}},\ and\ \citenamefont {et~al.}}]{2015arXiv150201589P}%
  \BibitemOpen
  \bibfield  {author} {\bibinfo {author} {\bibnamefont {{Planck
  Collaboration}}}, \bibinfo {author} {\bibfnamefont {P.~A.~R.}\ \bibnamefont
  {{Ade}}}, \bibinfo {author} {\bibfnamefont {N.}~\bibnamefont {{Aghanim}}},
  \bibinfo {author} {\bibfnamefont {M.}~\bibnamefont {{Arnaud}}}, \bibinfo
  {author} {\bibfnamefont {M.}~\bibnamefont {{Ashdown}}}, \bibinfo {author}
  {\bibfnamefont {J.}~\bibnamefont {{Aumont}}}, \bibinfo {author}
  {\bibfnamefont {C.}~\bibnamefont {{Baccigalupi}}}, \bibinfo {author}
  {\bibfnamefont {A.~J.}\ \bibnamefont {{Banday}}}, \bibinfo {author}
  {\bibfnamefont {R.~B.}\ \bibnamefont {{Barreiro}}}, \bibinfo {author}
  {\bibfnamefont {J.~G.}\ \bibnamefont {{Bartlett}}}, \ and\ \bibinfo {author}
  {\bibnamefont {et~al.}},\ }\href@noop {} {\bibfield  {journal} {\bibinfo
  {journal} {ArXiv e-prints}\ } (\bibinfo {year} {2015})},\ \Eprint
  {http://arxiv.org/abs/1502.01589} {arXiv:1502.01589 [astro-ph.CO]}
  \BibitemShut {NoStop}%
\bibitem [{\citenamefont {{Turner}}\ \emph {et~al.}(1993)\citenamefont
  {{Turner}}, \citenamefont {{White}},\ and\ \citenamefont
  {{Lidsey}}}]{1993PhRvD..48.4613T}%
  \BibitemOpen
  \bibfield  {author} {\bibinfo {author} {\bibfnamefont {M.~S.}\ \bibnamefont
  {{Turner}}}, \bibinfo {author} {\bibfnamefont {M.}~\bibnamefont {{White}}}, \
  and\ \bibinfo {author} {\bibfnamefont {J.~E.}\ \bibnamefont {{Lidsey}}},\
  }\href {\doibase 10.1103/PhysRevD.48.4613} {\bibfield  {journal} {\bibinfo
  {journal} {\prd}\ }\textbf {\bibinfo {volume} {48}},\ \bibinfo {pages} {4613}
  (\bibinfo {year} {1993})},\ \Eprint {http://arxiv.org/abs/astro-ph/9306029}
  {arXiv:astro-ph/9306029} \BibitemShut {NoStop}%
\bibitem [{\citenamefont {{Allen}}\ and\ \citenamefont
  {{Koranda}}(1994)}]{1994PhRvD..50.3713A}%
  \BibitemOpen
  \bibfield  {author} {\bibinfo {author} {\bibfnamefont {B.}~\bibnamefont
  {{Allen}}}\ and\ \bibinfo {author} {\bibfnamefont {S.}~\bibnamefont
  {{Koranda}}},\ }\href {\doibase 10.1103/PhysRevD.50.3713} {\bibfield
  {journal} {\bibinfo  {journal} {\prd}\ }\textbf {\bibinfo {volume} {50}},\
  \bibinfo {pages} {3713} (\bibinfo {year} {1994})},\ \Eprint
  {http://arxiv.org/abs/astro-ph/9404068} {arXiv:astro-ph/9404068} \BibitemShut
  {NoStop}%
\bibitem [{\citenamefont {{Ng}}\ and\ \citenamefont
  {{Speliotopoulos}}(1995)}]{1995PhRvD..52.2112N}%
  \BibitemOpen
  \bibfield  {author} {\bibinfo {author} {\bibfnamefont {K.-W.}\ \bibnamefont
  {{Ng}}}\ and\ \bibinfo {author} {\bibfnamefont {A.~D.}\ \bibnamefont
  {{Speliotopoulos}}},\ }\href {\doibase 10.1103/PhysRevD.52.2112} {\bibfield
  {journal} {\bibinfo  {journal} {\prd}\ }\textbf {\bibinfo {volume} {52}},\
  \bibinfo {pages} {2112} (\bibinfo {year} {1995})},\ \Eprint
  {http://arxiv.org/abs/astro-ph/9405043} {arXiv:astro-ph/9405043} \BibitemShut
  {NoStop}%
\bibitem [{\citenamefont {{Wang}}(1996)}]{1996PhRvD..53..639W}%
  \BibitemOpen
  \bibfield  {author} {\bibinfo {author} {\bibfnamefont {Y.}~\bibnamefont
  {{Wang}}},\ }\href {\doibase 10.1103/PhysRevD.53.639} {\bibfield  {journal}
  {\bibinfo  {journal} {\prd}\ }\textbf {\bibinfo {volume} {53}},\ \bibinfo
  {pages} {639} (\bibinfo {year} {1996})},\ \Eprint
  {http://arxiv.org/abs/astro-ph/9501116} {arXiv:astro-ph/9501116} \BibitemShut
  {NoStop}%
\bibitem [{\citenamefont {{Weinberg}}(2004)}]{2004PhRvD..69b3503W}%
  \BibitemOpen
  \bibfield  {author} {\bibinfo {author} {\bibfnamefont {S.}~\bibnamefont
  {{Weinberg}}},\ }\href {\doibase 10.1103/PhysRevD.69.023503} {\bibfield
  {journal} {\bibinfo  {journal} {\prd}\ }\textbf {\bibinfo {volume} {69}},\
  \bibinfo {eid} {023503} (\bibinfo {year} {2004})},\ \Eprint
  {http://arxiv.org/abs/astro-ph/0306304} {arXiv:astro-ph/0306304 [astro-ph]}
  \BibitemShut {NoStop}%
\bibitem [{\citenamefont {{Pritchard}}\ and\ \citenamefont
  {{Kamionkowski}}(2005)}]{2005AnPhy.318....2P}%
  \BibitemOpen
  \bibfield  {author} {\bibinfo {author} {\bibfnamefont {J.~R.}\ \bibnamefont
  {{Pritchard}}}\ and\ \bibinfo {author} {\bibfnamefont {M.}~\bibnamefont
  {{Kamionkowski}}},\ }\href {\doibase 10.1016/j.aop.2005.03.005} {\bibfield
  {journal} {\bibinfo  {journal} {Annals of Physics}\ }\textbf {\bibinfo
  {volume} {318}},\ \bibinfo {pages} {2} (\bibinfo {year} {2005})},\ \Eprint
  {http://arxiv.org/abs/astro-ph/0412581} {arXiv:astro-ph/0412581} \BibitemShut
  {NoStop}%
\bibitem [{\citenamefont {{Bashinsky}}(2005)}]{2005astro.ph..5502B}%
  \BibitemOpen
  \bibfield  {author} {\bibinfo {author} {\bibfnamefont {S.}~\bibnamefont
  {{Bashinsky}}},\ }\href@noop {} {\bibfield  {journal} {\bibinfo  {journal}
  {ArXiv Astrophysics e-prints}\ } (\bibinfo {year} {2005})},\ \Eprint
  {http://arxiv.org/abs/astro-ph/0505502} {arXiv:astro-ph/0505502} \BibitemShut
  {NoStop}%
\bibitem [{\citenamefont {{Weinberg}}(2008)}]{2008cosm.book.....W}%
  \BibitemOpen
  \bibfield  {author} {\bibinfo {author} {\bibfnamefont {S.}~\bibnamefont
  {{Weinberg}}},\ }\href@noop {} {\emph {\bibinfo {title} {Cosmology, by Steven
  Weinberg.~ISBN 978-0-19-852682-7.~Published by Oxford University Press,
  Oxford, UK, 2008.}}}\ (\bibinfo  {publisher} {Oxford University Press},\
  \bibinfo {year} {2008})\BibitemShut {NoStop}%
\bibitem [{\citenamefont {{Stefanek}}\ and\ \citenamefont
  {{Repko}}(2013)}]{2013PhRvD..88h3536S}%
  \BibitemOpen
  \bibfield  {author} {\bibinfo {author} {\bibfnamefont {B.~A.}\ \bibnamefont
  {{Stefanek}}}\ and\ \bibinfo {author} {\bibfnamefont {W.~W.}\ \bibnamefont
  {{Repko}}},\ }\href {\doibase 10.1103/PhysRevD.88.083536} {\bibfield
  {journal} {\bibinfo  {journal} {\prd}\ }\textbf {\bibinfo {volume} {88}},\
  \bibinfo {eid} {083536} (\bibinfo {year} {2013})},\ \Eprint
  {http://arxiv.org/abs/1207.7285} {arXiv:1207.7285 [hep-ph]} \BibitemShut
  {NoStop}%
\bibitem [{\citenamefont {Rebhan}\ and\ \citenamefont
  {Schwarz}(1994)}]{PhysRevD.50.2541}%
  \BibitemOpen
  \bibfield  {author} {\bibinfo {author} {\bibfnamefont {A.~K.}\ \bibnamefont
  {Rebhan}}\ and\ \bibinfo {author} {\bibfnamefont {D.~J.}\ \bibnamefont
  {Schwarz}},\ }\href {\doibase 10.1103/PhysRevD.50.2541} {\bibfield  {journal}
  {\bibinfo  {journal} {Phys. Rev. D}\ }\textbf {\bibinfo {volume} {50}},\
  \bibinfo {pages} {2541} (\bibinfo {year} {1994})}\BibitemShut {NoStop}%
\bibitem [{\citenamefont {Feshchenko}\ \emph {et~al.}(1966)\citenamefont
  {Feshchenko}, \citenamefont {Shkil},\ and\ \citenamefont {Nikolenko}}]{FSN}%
  \BibitemOpen
  \bibfield  {author} {\bibinfo {author} {\bibfnamefont {S.~F.}\ \bibnamefont
  {Feshchenko}}, \bibinfo {author} {\bibfnamefont {N.~I.}\ \bibnamefont
  {Shkil}}, \ and\ \bibinfo {author} {\bibfnamefont {L.~D.}\ \bibnamefont
  {Nikolenko}},\ }\href {https://archive.org/details/nasa_techdoc_19670019838}
  {\emph {\bibinfo {title} {Asymptotic methods in the theory of linear
  differential equations, S.F. Feshchenko, N.I. Shkil, and L.D. Nikolenko.
  Translated by Scripta Technica}}}\ (\bibinfo  {publisher} {American Elsevier
  Pub. Co New York},\ \bibinfo {year} {1966})\ pp.\ \bibinfo {pages} {xvi, 270
  p.},\ \bibinfo {note} {translated for the National Aeronautics and Space
  Administration by John F. Holman and Co. Inc. NASw-1495 20546, available at
  https://archive.org/details/nasa\_ techdoc\_19670019838 online}\BibitemShut
  {NoStop}%
\bibitem [{\citenamefont {Wren}\ and\ \citenamefont {Malik}(2017)}]{DPS}%
  \BibitemOpen
  \bibfield  {author} {\bibinfo {author} {\bibfnamefont {A.~J.}\ \bibnamefont
  {Wren}}\ and\ \bibinfo {author} {\bibfnamefont {K.~A.}\ \bibnamefont
  {Malik}},\ }\href {\doibase 10.1103/PhysRevD.95.083526} {\bibfield  {journal}
  {\bibinfo  {journal} {Phys. Rev.}\ }\textbf {\bibinfo {volume} {D95}},\
  \bibinfo {pages} {083526} (\bibinfo {year} {2017})},\ \Eprint
  {http://arxiv.org/abs/1603.07577} {arXiv:1603.07577 [gr-qc]} \BibitemShut
  {NoStop}%
\bibitem [{\citenamefont {{Wren}}\ and\ \citenamefont
  {{Malik}}(2016)}]{GWGithub}%
  \BibitemOpen
  \bibfield  {author} {\bibinfo {author} {\bibfnamefont {A.~J.}\ \bibnamefont
  {{Wren}}}\ and\ \bibinfo {author} {\bibfnamefont {K.~A.}\ \bibnamefont
  {{Malik}}},\ }\href {TBC} {\enquote {\bibinfo {title} {{Mathematica notebooks
  for \emph{TBC - THIS PAPER}}},}\ } (\bibinfo {year} {2016}),\ \bibinfo {note}
  {{available at TBC\ }}\BibitemShut {NoStop}%
\bibitem [{\citenamefont {{Padmanabhan}}(2006)}]{2006AIPC..843..111P}%
  \BibitemOpen
  \bibfield  {author} {\bibinfo {author} {\bibfnamefont {T.}~\bibnamefont
  {{Padmanabhan}}},\ }in\ \href {\doibase 10.1063/1.2219327} {\emph {\bibinfo
  {booktitle} {Graduate School in Astronomy: X}}},\ \bibinfo {series} {American
  Institute of Physics Conference Series}, Vol.\ \bibinfo {volume} {843},\
  \bibinfo {editor} {edited by\ \bibinfo {editor} {\bibfnamefont
  {S.}~\bibnamefont {{Daflon}}}, \bibinfo {editor} {\bibfnamefont
  {J.}~\bibnamefont {{Alcaniz}}}, \bibinfo {editor} {\bibfnamefont
  {E.}~\bibnamefont {{Telles}}}, \ and\ \bibinfo {editor} {\bibfnamefont
  {R.}~\bibnamefont {{de la Reza}}}}\ (\bibinfo {year} {2006})\ pp.\ \bibinfo
  {pages} {111--166},\ \Eprint {http://arxiv.org/abs/astro-ph/0602117}
  {arXiv:astro-ph/0602117} \BibitemShut {NoStop}%
\bibitem [{\citenamefont {{Wright}}(2006)}]{2006PASP..118.1711W}%
  \BibitemOpen
  \bibfield  {author} {\bibinfo {author} {\bibfnamefont {E.~L.}\ \bibnamefont
  {{Wright}}},\ }\href {\doibase 10.1086/510102} {\bibfield  {journal}
  {\bibinfo  {journal} {\pasp}\ }\textbf {\bibinfo {volume} {118}},\ \bibinfo
  {pages} {1711} (\bibinfo {year} {2006})},\ \Eprint
  {http://arxiv.org/abs/astro-ph/0609593} {arXiv:astro-ph/0609593} \BibitemShut
  {NoStop}%
\bibitem [{\citenamefont {{Malik}}\ and\ \citenamefont
  {{Wands}}(2009)}]{2009PhR...475....1M}%
  \BibitemOpen
  \bibfield  {author} {\bibinfo {author} {\bibfnamefont {K.~A.}\ \bibnamefont
  {{Malik}}}\ and\ \bibinfo {author} {\bibfnamefont {D.}~\bibnamefont
  {{Wands}}},\ }\href {\doibase 10.1016/j.physrep.2009.03.001} {\bibfield
  {journal} {\bibinfo  {journal} {\physrep}\ }\textbf {\bibinfo {volume}
  {475}},\ \bibinfo {pages} {1} (\bibinfo {year} {2009})},\ \Eprint
  {http://arxiv.org/abs/0809.4944} {arXiv:0809.4944 [astro-ph]} \BibitemShut
  {NoStop}%
\bibitem [{\citenamefont {{Ryden}}(2003)}]{2003itc..book.....R}%
  \BibitemOpen
  \bibfield  {author} {\bibinfo {author} {\bibfnamefont {B.}~\bibnamefont
  {{Ryden}}},\ }\href@noop {} {\emph {\bibinfo {title} {Introduction to
  cosmology / Barbara Ryden.~San Francisco, CA, USA: Addison Wesley, ISBN
  0-8053-8912-1, 2003, IX + 244 pp.}}}\ (\bibinfo  {publisher} {Addison
  Wesley},\ \bibinfo {year} {2003})\BibitemShut {NoStop}%
\bibitem [{\citenamefont {{Lyth}}\ and\ \citenamefont
  {{Liddle}}(2009)}]{2009pdp..book.....L}%
  \BibitemOpen
  \bibfield  {author} {\bibinfo {author} {\bibfnamefont {D.~H.}\ \bibnamefont
  {{Lyth}}}\ and\ \bibinfo {author} {\bibfnamefont {A.~R.}\ \bibnamefont
  {{Liddle}}},\ }\href@noop {} {\emph {\bibinfo {title} {The Primordial Density
  Perturbation, by David H.~Lyth, Andrew R.~Liddle, Cambridge, UK: Cambridge
  University Press, 2009}}}\ (\bibinfo {year} {2009})\BibitemShut {NoStop}%
\bibitem [{\citenamefont {{Carrilho}}\ and\ \citenamefont
  {{Malik}}(2016)}]{2016JCAP...02..021C}%
  \BibitemOpen
  \bibfield  {author} {\bibinfo {author} {\bibfnamefont {P.}~\bibnamefont
  {{Carrilho}}}\ and\ \bibinfo {author} {\bibfnamefont {K.~A.}\ \bibnamefont
  {{Malik}}},\ }\href {\doibase 10.1088/1475-7516/2016/02/021} {\bibfield
  {journal} {\bibinfo  {journal} {\jcap}\ }\textbf {\bibinfo {volume} {2}},\
  \bibinfo {eid} {021} (\bibinfo {year} {2016})},\ \Eprint
  {http://arxiv.org/abs/1507.06922} {arXiv:1507.06922 [astro-ph.CO]}
  \BibitemShut {NoStop}%
\end{thebibliography}

\appendix \section{Calculation of the matching conditions for approximating primordial gravitational waves}
\label{sec:calculation-of-the-matching-conditions-for-approximating-primordial-gravitational-waves}

In this appendix, we calculate the matching conditions needed to obtain  the solution in \eref{eq:match} of Section~\ref{sec:modelling-gravitational-waves-through-the-whole-radiation--matter-epoch}.  We start by matching the values of the functions $\hr$ of \eref{eq:flat-rad-only-soln3} and $h_{(2)}$ of \ref{eq:second-order-trig2} at the conformal time $\tau=\infrac{1}{k}.$ This gives
\bas \label{eq:fval}
\sin(1)&=\frac{B_\text{s}\sin \left(\lambda\left(k,\frac{1}{k}\right)\right)+B_\text{c}\cos \left(\lambda\left(k,\frac{1}{k}\right)\right)}{\frac{1}{k}  (\frac{1}{k} +4)}\\
&=
\frac{k^2}{4k+1}
\left\lbrace B_\text{s}\sin\left(\lambda\left(k,\frac{1}{k}\right)\right) +B_\text{c}\cos\left(\lambda\left(k,\frac{1}{k}\right)\right)\right\rbrace,
\eas
where
\be \label{eq:cts}
\lambda\left(k,\frac{1}{k}\right)=1+\frac{1}{4 k}\ln \left(1+4k\right).
\ee

Note that
\be \label{eq:d1}
\frac{\partial \left(\lambda(k,\tau)\right)}{\partial \tau}\bigg|_{\tau=\frac{1}{k}}
=k-\frac{1}{k \tau ^2+4 k \tau }\bigg|_{\tau=\frac{1}{k}}
=k-\frac{k}{4k+1}
=\frac{4k^2}{4k+1},
\ee
and 
\be \label{eq:d2}
\frac{\partial \Big(\frac{1}{\tau\left(\tau+4\right)}\Big)}{\partial \tau}\Bigg|_{\tau=\frac{1}{k}}
=-\frac{2 (\tau +2)}{\tau ^2 (\tau +4)^2}\Bigg|_{\tau=\frac{1}{k}}
=-\frac{2 k^3 (2 k+1)}{(4 k+1)^2}.
\ee 
Using  Eqs.~\ref{eq:d1} and \ref{eq:d2} to calculate $h'_{(2)}\left(\infrac{1}{k}\right),$ we match the first derivatives of with respect to $\tau$ of $\hr$ and $h_{(2)}$ at $\tau=\infrac{1}{k}$ to get
\bml \label{eq:fderiv}
k\cos(1)-k\sin(1)
=
\frac{4k^4}{\left(4k+1\right)^2}
\Bigg\lbrace B_\text{s}\cos\left(\lambda\left(k,\frac{1}{k}\right)\right)-B_\text{c}\sin\left(\lambda\left(k,\frac{1}{k}\right)\right)\Bigg\rbrace \\
-\frac{2 k^3 (2 k+1)}{(4 k+1)^2}\Bigg\lbrace B_\text{s}\sin\left(\lambda\left(k,\frac{1}{k}\right)\right)+B_\text{c}\cos\left(\lambda\left(k,\frac{1}{k}\right)\right)\Bigg\rbrace.
\eml\\

We can solve Eqs.~\ref{eq:fval} and \ref{eq:fderiv} simultaneously to find
\bas \label{eq:b-s}
B_\text{s}
&=
\frac{4 k+1}{4 k^3} \left\lbrace 4 k\left[
\cos(1)\cos \left(\lambda\left(k,\frac{1}{k}\right)\right) +\sin (1)\sin\left(\lambda\left(k,\frac{1}{k}\right)\right)\right]+\left[\sin(1)+\cos (1)\right]\cos\left(\lambda\left(k,\frac{1}{k}\right)\right)\right\rbrace\\
&=
\frac{4 k+1}{4 k^3} \left\lbrace 4 k\cos \left(1-\lambda\left(k,\frac{1}{k}\right)\right) +\left[\sin(1)+\cos (1)\right]\cos\left(\lambda\left(k,\frac{1}{k}\right)\right)\right\rbrace\\
&=
\frac{4 k+1}{4 k^3} \left\lbrace 4 k\cos \left(\frac{1}{4k}\ln\left(1+4k\right)\right) +\left[\sin(1)+\cos (1)\right]\cos\left(1+\frac{1}{4k}\ln\left(1+4k\right)\right)\right\rbrace
\eas
and
\bas  \label{eq:b-c}
B_\text{c}
&=
\frac{4 k+1}{4 k^3} \left\lbrace 4 k\left[
\sin(1)\cos \left(\lambda\left(k,\frac{1}{k}\right)\right) -\cos (1)\sin\left(\lambda\left(k,\frac{1}{k}\right)\right)\right]-\left[\sin(1)+\cos (1)\right]\sin\left(\lambda\left(k,\frac{1}{k}\right)\right)\right\rbrace\\
&=
\frac{4 k+1}{4 k^3} \left\lbrace 4 k\sin \left(1-\lambda\left(k,\frac{1}{k}\right)\right) -\left[\sin(1)+\cos (1)\right]\sin\left(\lambda\left(k,\frac{1}{k}\right)\right)\right\rbrace\\
&=
\frac{4 k+1}{4 k^3} \left\lbrace - 4 k\sin \left(\frac{1}{4k}\ln\left(1+4k\right)\right) -\left[\sin(1)+\cos (1)\right]\sin\left(1+\frac{1}{4k}\ln\left(1+4k\right)\right)\right\rbrace,
\eas
where we recalled the definition of $\lambda(k,\tau)$ from \eref{eq:lamdba}.
In going from the first to second lines of Eqs.~\ref{eq:b-s} and \ref{eq:b-c}, we have used standard trigonometric addition formulae.\\

From \eref{eq:second-order-trig2}, this gives us
\begin{subequations}
\begin{align} 
\begin{split}\label{eq:trigs-a}
h_{(2)}(\tau)
&=
\frac{4 k+1}{4 k^3 \tau  (4+\tau)}
\left[
4k\left\lbrace
\cos\left(\frac{1}{4k}\ln\left(1+4k\right)\right)\sin \left(\lambda(k,t)\right)
-
\sin\left(\frac{1}{4k}\ln\left(1+4k\right)\right)\cos \left(\lambda(k,t)\right)
\right\rbrace\right.
\\
&\left.\qquad+\big[\sin (1)+\cos (1)\big]
\bigg\lbrace\cos\left(1+\frac{1}{4k}\ln\left(1+4k\right)\right)\sin \left(\lambda(k,t)\right)\right.
\\
&\qquad\qquad\qquad\qquad\qquad\qquad\left.
-\sin\left(1+\frac{1}{4k}\ln\left(1+4k\right)\right)\cos \left(\lambda(k,t)\right)\bigg\rbrace
\right]
\end{split}
\\[9pt]
\begin{split}\label{eq:trigs-b}
&=
\frac{4 k+1}{4 k^3 \tau  (4+\tau)}
\bigg[
4k \sin \Big(k\tau + L(k,t)\Big)+\left[\sin (1)+\cos (1)\right]
\sin \Big(k\tau + L(k,t)-1\Big)
\bigg],
\end{split}
\end{align}
\end{subequations}
where again we recalled the definition of $\lambda(k,\tau)$ from \eref{eq:lamdba}, and we have defined
\be
L(k,\tau)=\frac{1}{4 k}\ln \left(\frac{1+4\,\tau^{-1} }{1+4\,k\hfill}\right).
\ee
We used standard trigonometric addition formulae to go from \eref{eq:trigs-a} to \eref{eq:trigs-b}.\\

From \eref{eq:flat-rad-only-soln3} and \eref{eq:trigs-b}, the matching approximation is therefore
\begin{equation}\label{eq:match-a}
\hmatch(\tau)=
\begin{dcases}
\frac{\sin(k\tau)}{k\tau}
&\mbox{if } \tau\le\frac{1}{k}
\\[9pt]
\frac{4 k+1}{4 k^3 \tau  (4+\tau)}
\bigg[
\,4 k\, \sin \Big(k\tau + L(k,t)\Big)
+ \mu
\sin \Big(k\tau + L(k,\tau)-1\Big)
\bigg]
 &\mbox{if } \tau\ge\frac{1}{k},
\end{dcases}
\end{equation}
where $
\mu
=
\sin\left(1\right)+\cos\left(1\right)
=
1.38177...\, .
$

\section{Calculation of the anisotropic stress solution}
\label{sec:calculation-of-the-anisotropic-stress-solution}

The method described in \rfcite{DPS} works best for equations like
\ba
\mathbf{A} \mathbf{f}''(\tau) + \mathbf{C} \mathbf{f}'(\tau) + \mathbf{B} \mathbf{f}(\tau) = \mathbf{0}. 
\ea
so we take \eref{eq:gov-tau-stress} and write it as a system of ordinary second order differential equations to get\footnote{\textbf{Note.} We are neglecting the term $\int_{0}^{\tau}K'(\tau-t)h'(t)dt$ that arises from differentiating~\eref{eq:gov-tau-stress} because numerically the value is \textit{negligible}.}
\ba
\mathbf{A} \mathbf{X}''(\tau) + \mathbf{C} \mathbf{X}'(\tau) + \mathbf{B} \mathbf{X}(\tau) = \mathbf{0}. 
\ea
where
\[
   \mathbf{X}(\tau)=
  \left[ {\begin{array}{cc}
   x_{1}(\tau) \\
   x_{2}(\tau) \\
  \end{array} } \right] =
    \left[ {\begin{array}{cc}
   h(\tau) \\
   h'(\tau) \\
  \end{array} } \right],
\]
and 
\[
   \mathbf{A}=
  \left[ {\begin{array}{cc}
   1 & 0\\
   0 & 1\\
  \end{array} } \right], \quad
     \mathbf{C}(\tau)=
  \left[ {\begin{array}{cc}
   0 & -1\\
   0 & f_{2}(\tau)/f_{1}(\tau)\\
  \end{array} } \right], \quad
     \mathbf{B}(\tau)=
  \left[ {\begin{array}{cc}
   0 & 0\\
   f_{4}(\tau)/f_{1}(\tau) & f_{3}(\tau)/f_{1}(\tau)\\
  \end{array} } \right] ,
\]
where we have defined
\begin{align}
f_{1}(\tau) &= -\frac{1}{96 f_{\nu}(0)}\frac{\tau^{2}(1+\tau)(4+\tau)^{2}}{(2+\tau)^{2}}, \\
f_{2}(\tau) &= -\frac{1}{96 f_{\nu}(0)}\frac{\tau(4+\tau)\{32+\tau(64+\tau[36+7\tau])\}}{(2+\tau)^{3}}, \\
f_{3}(\tau) &= -\frac{1}{96 f_{\nu}(0)}\frac{32+\tau\left\{80+\tau[44+8\tau+k^{2}(1+\tau)(4+\tau)^{2}]\right\}}{(2+\tau)^{2}} - \frac{1}{15}, \\
f_{4}(\tau) &= -\frac{1}{96 f_{\nu}(0)}\frac{k^{2} \tau \left\{4+\tau [16+\tau (16+3\tau) ]\right\}}{121569(2+\tau)^{3}}.
\end{align}
Following \rfcite{DPS} we need to further decompose $\mathbf{B}(\tau) = \mathbf{B_{0}}(\tau) + \mathbf{B_{-2}}(\tau) k^{2}$ and those matrices are
\[
     \mathbf{B_{0}}(\tau)=
  \left[ {\begin{array}{cc}
   0 & 0\\
   0 & b_{0}(\tau)\
  \end{array} } \right], \quad \text{and} \quad
     \mathbf{B_{-2}}(\tau)=
  \left[ {\begin{array}{cc}
   0 & 0\\
   b_{-2}(\tau)& 1\\
  \end{array} } \right] .
\]
where 
\begin{align}
b_{0}(\tau) &= \frac{4\left\{ 8f_{\nu}(0)(2+\tau)^{2} + 5 (8+\tau[20+\tau(11+2\tau)]) \right\}}{5 \tau^{2}(1+\tau)(4+\tau)^{2}}, \\
b_{-2}(\tau) &= \frac{2}{\tau} + \frac{1}{1+\tau} - \frac{2}{2+\tau} + \frac{2}{4+\tau}. 
\end{align}

Finally, we use the \textit{Double Power Series Method} in \rfcite{GWGithub} to get Eqs.~(\ref{eq:flat-rad-only-soln-stress}) and (\ref{eq:third-order-stress}).

\end{document}